\documentclass[lettersize,journal]{IEEEtran}
\usepackage{amsmath,amsfonts}
\usepackage{algorithmic}
\usepackage{array}
\usepackage{textcomp}
\usepackage{stfloats}
\usepackage{url}
\usepackage{verbatim}
\usepackage{graphicx}
\def\BibTeX{{\rm B\kern-.05em{\sc i\kern-.025em b}\kern-.08em
		T\kern-.1667em\lower.7ex\hbox{E}\kern-.125emX}}
\usepackage{balance}

\usepackage[numbers]{natbib}
\usepackage{hyperref}
\usepackage{tabularx}
\usepackage{booktabs} \usepackage{makecell}
\usepackage{subcaption}

\newcolumntype{g}{X} \newcolumntype{s}{>{\hsize=.5\hsize}X}              \newcolumntype{Y}{>{\centering\arraybackslash}X}

\begin{document}
\title{An Investigation of the Relation Between\\Immersion and Learning Across Three Domains}
	\author{Paolo Boffi
		\and 
		Alberto Gallace
		\and 
		Pier Luca Lanzi
		\thanks{Paolo Boffi (paolo.boffi@polimi.it) is with the Department of Electronics, Informatics, and Bioengineering (DEIB) at Polytechnic University of Milan and with the Mind and Behavior Technological Center (MiBTec) at University of Milan-Bicocca. Alberto Gallace (alberto.gallace1@unimib.it) is with the Mind and Behavior Technological Center (MiBTec) at University of Milan-Bicocca. Pier Luca Lanzi (pierluca.lanzi@polimi.it) is with the Department of Electronics, Informatics, and Bioengineering (DEIB) at Polytechnic University of Milan. Contact author: Paolo Boffi (paolo.boffi@polimi.it).}	
	}
		
	\maketitle
	
	\begin{abstract}
		We investigate the relationship between immersion and learning across three domains (cultural heritage, environmental awareness, and high school physics) through the lens of the  Cognitive Affective Model of Immersive Learning (CAMIL) framework. We present three applications we developed for this investigation, highlighting their shared design elements and domain-specific mechanics. Using a common evaluation protocol across lab studies and a classroom deployment, we assessed learning outcomes, user experience, technology acceptance, presence/embodiment, and cybersickness. Our results show that immersive virtual reality led to higher scores for presence, user experience, and technology acceptance. In contrast, learning outcomes were mixed. In immediate post-test evaluations, factual knowledge scores were comparable between immersive virtual reality and control groups. In the end, we synthesize design guidelines that outline when immersive virtual reality might be most beneficial in didactic contexts, and we provide CAMIL-informed recommendations and strategies to improve learning outcomes and overall experiential quality.

 	\end{abstract}
	
	
\section{Introduction}
\label{sec:introduction}
Virtual Reality (VR) has rapidly grown over the past decade as head-mounted displays have become more affordable and less intrusive~\cite{jensen-konradsen2018, radianti2020}. In education, VR promises unique affordances: it allows users to visualize abstract or invisible phenomena, explore lost or unreachable places, and rehearse risky procedures safely~\cite{radianti2020, dinatale2020, buttussi2018}. It also affords tightly controlled, instrumented environments that enable precise measurement of behavior and learning~\cite{dinatale2020, jensen-konradsen2018, buttussi2018}. Nevertheless, the evidence for high-immersion VR improving learning remains mixed: reviews report gains in presence and engagement without consistent advantages on test performance, and in some cases, high immersion can hinder outcomes~\cite{merchant2014, radianti2020, pellas2021, wu2020, makransky2019b}. 
All previously published studies have been conducted on single applications and, while they can be qualitatively compared with other studies, quantitative comparisons are infeasible due to differences in experimental protocols. In this paper, we investigate the relation between immersion and learning using the same lens, the Cognitive Affective Model of Immersive Learning (CAMIL) framework \cite{makransky2021camil}, and a common experimental protocol applied to three application domains (cultural heritage, environmental awareness, and high school physics).

A key challenge is isolating immersion's specific contribution to learning from other design variables -- such as interaction modality, locomotion, or representational fidelity -- that also shape experience via presence and agency. CAMIL frames these effects as operating through psychological routes (presence, agency) that in turn influence interest, motivation, cognitive load, and related constructs~\cite{makransky2021camil}. In this paper, CAMIL is used as a lens to interpret immersion's impact on learning; the goal is not to systematically test interactions among CAMIL's technological factors, but to examine how differing levels of immersion relate to outcomes when other elements are held constant or matched as closely as feasible.

This paper investigates immersion's role across three educational domains -- cultural heritage, environmental awareness, and physics -- using a shared evaluation backbone and the same theoretical lens. The studies comprise lab experiments and an ecological classroom deployment to balance control and external validity. The contributions are: a cross-domain evidence disentangling immersion from control factors using a consistent measurement strategy; a CAMIL-based analysis that connects presence and agency to motivational and affective outcomes alongside learning; and actionable design recommendations for when and how to deploy Immersive Virtual Reality (IVR) in instructional settings.

\section{Related Work}
\label{sec:related_work}
\subsection{Virtual Reality: Definitions and Main Characteristics}
\label{subsec:vr-characteristics}

VR encompasses interactive, computer-generated environments that users can explore and act within through multisensory stimulation and natural interaction. It is hard to find an univocal definition in the literature~\cite{ambrosiofidalgo2020}.~\citet{biocca1995} (as cited in~\cite[p.~2]{radianti2020}) described it as ``the sum of the hardware and software systems that seek to perfect an all-inclusive, immersive, sensory illusion of being present in another environment''.~\citet{chavez2018} offered a slightly different perspective, defining VR as a graphically rendered, three-dimensional digital model of an artificial environment designed to give the user a feeling of being immersed in a given place through multisensory stimulation. Lee and Wong emphasized interaction, defining VR as ``a way of simulating or replicating an environment that can be explored and interacted with by a person''~\cite[p.~50]{lee2014}. 

From the previous definitions, we can infer some key characteristics of VR.~\citet{biocca1995} emphasize the collaboration between hardware and software systems to induce \emph{immersion} and the illusion of being \emph{present} within the simulated environment -- concepts also highlighted by~\citet{chavez2018}. The latter additionally stress the importance of \emph{multisensory feedback}, which can enhance the immersive experience. Finally,~\citet{lee2014} place the user at the center of the VR experience, emphasizing the significance of enabling \emph{interactions} within the environment. 

Based on these insights, we can identify four fundamental features that a VR system should provide: \emph{immersion}, \emph{presence}, \emph{interactivity}, and \emph{multisensoriality}~\citep{walsh2002, ryan2015}. \emph{Interactivity} refers to the degree to which a user can modify the VR environment in real-time~\citep{steuer1995, johnson2022}. \emph{Multisensoriality} instead refers to integrating multiple sensory modalities -- such as visual, auditory, and haptic feedback -- to create a more realistic and engaging experience for the user~\cite{velasco2021}.

Immersion and presence (often referred to as the \emph{Sense of Presence}) are sometimes treated as interchangeable~\citep{skarbez2017,jensen-konradsen2018}, yet they capture distinct dimensions of the VR experience. There is still a debate in the literature on these two concepts, with two main definitions. In Witmer and Singer's account, immersion is a psychological state in which users feel enveloped by, included in, and able to act within the Virtual Environment (VE)~\citep{witmer1998,nilsson2016}. They identified a positive relationship between immersion and presence, defined as ``the psychological state of `being there' when an environment engages the senses, captures attention, and supports active involvement''~\cite[p.~298]{witmer2005}. Presence, on this view, is inherently subjective and shaped by prior experience, expectations, and cognitive abilities. Slater offers a complementary, technological perspective: immersion is an objective property of a VR system, fixed by its display, tracking, and interaction capabilities~\citep{slater1994,slater2003,slater2018}. Individual users then respond to a given level of immersion with different degrees of presence, which Slater characterizes as the feeling of being in the VE, the VE's dominance over the physical world as the user's primary frame of reference, and experiencing the VE as a \emph{place visited} rather than a set of rendered images~\cite{slater2003}.

Following Slater's definition, VR systems have been categorized based on their immersion degree -- namely, how much of the real world they can hide from the user's sight~\cite{jensen-konradsen2018, cipresso2018, dinatale2020}. Following this criterion, three categories have been proposed: non-immersive (desktop monitor, mouse/keyboard/gamepad), semi-immersive (large screens/projection plus advanced input), and immersive HMD-based systems affording first-person, head/body-tracked interaction. 

In this work, we will refer to non-immersive systems as \emph{Desktop VR} (DVR) and to immersive systems as \emph{Immersive VR} (IVR); we use \emph{VR} for the technology in general.

\subsection{Educational Virtual Reality}
\label{subsec:educational-vr}

VR is widely recognized as a powerful educational tool, enabling the visualization of abstract concepts, safe rehearsal of procedures in risk-free settings, and exploration of otherwise inaccessible environments~\citep{radianti2020, kaminska2019, jensen-konradsen2018, ambrosiofidalgo2020}. These affordances align with established learning paradigms that have been applied to (or adapted for) VR. In fact, it is crucial to understand how humans learn and interact with technology to provide an optimal combination of traditional and novel educational approaches. This section reviews some of the most commonly cited learning theories and paradigms and their relationship with VR as a novel educational tool.

\emph{Constructivism.} According to this theory, learners are not passive receivers but active \emph{constructors} of knowledge. They predict the outcomes of their actions based on existing mental models, act within the environment, and then receive feedback. If the outcome matches expectations, their model is reinforced; if not, it is updated~\cite{christou2010}. VR is particularly well-suited for constructivist learning activities, as it allows users to actively interact with the environment and receive multisensory feedback from their actions~\citep{radianti2020,christou2010}. Multisensory input further strengthens the learning process, providing multiple channels for learners to receive feedback and update their understanding. In this sense, VR's capacity to create immersive, interactive, and adaptable environments is a strong resource for constructivism-based learning~\citep{dinatale2020, kaminska2019}.

\emph{Embodied Cognition.} This theory states that learning is not only a mental process but involves the entire body and sensory-motor activities~\citep{wilson2002, ionescu2014}. Since VR engages multiple senses and enables full-body interaction within the environment, it offers strong potential for embodied learning experiences. Additionally, integrating VR experiences with embodied avatars makes users perceive the interaction with the environment even more, potentially leading to improved learning results~\citep{lindgren2013, johnson-glenberg2021}.

\emph{Cognitive Theory of Multimedia Learning (CTML).} This framework provides insights into how learners process and integrate multimedia information based on three principles: first, humans process information through two channels -- visual and auditory; second, those channels have limited processing capacity; third, meaningful learning occurs through active cognitive processing, involving the selection, organization, and integration of information with prior knowledge~\cite{mayer2014}. VR can be a powerful tool for multimedia learning within the context of CTML theory, as it combines multiple modalities -- visual, auditory, and haptic -- coherently, enhancing the learning experience~\cite{radianti2020}.  

\begin{figure}[t]
	\centering
	\includegraphics[width=\linewidth]{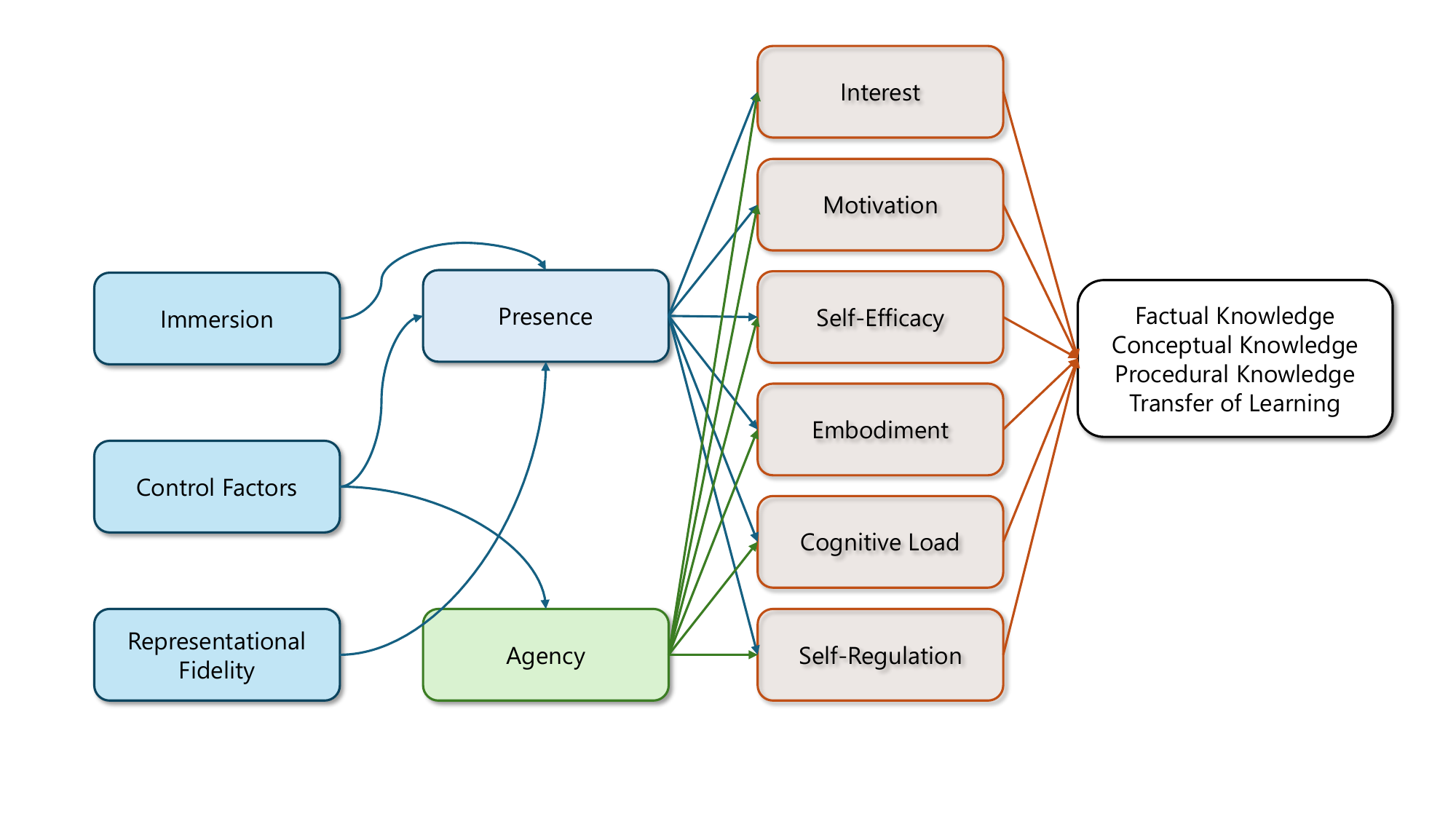}
	\caption{CAMIL's framework, adapted from~\citep{makransky2021camil}.}
	\label{fig:camil}
\end{figure}

\emph{The Cognitive Affective Model of Immersive Learning (CAMIL)}. While the theories above describe \emph{how} learners process information, they do not explicitly model how \emph{VR-specific} technological affordances influence learning. The Cognitive Affective Model of Immersive Learning (CAMIL; \autoref{fig:camil})~\cite{makransky2021camil} addresses this gap by linking core technological factors -- \emph{immersion}, \emph{control factors} (interaction design, immediacy, input modality), and \emph{representational fidelity} -- to two key psychological affordances: \emph{presence} (being there) and \emph{agency} (feeling of generating/controlling actions). 

CAMIL posits that presence and agency shape a set of cognitive-affective variables relevant to learning: \emph{interest}, \emph{motivation}, \emph{self-efficacy}, \emph{embodiment}, \emph{cognitive load}, and \emph{self-regulation}. These variables, in turn, are associated with learning outcomes that can be categorized using Bloom's taxonomy~\cite{anderson2001} -- namely, \emph{factual} knowledge (recall of discrete terms and elements), \emph{conceptual} knowledge (understanding relationships and principles), \emph{procedural} knowledge (knowing how and when to execute methods), and \emph{transfer} (applying what was learned to novel problems and contexts). In this work, CAMIL thus serves as a VR-specific, mechanistic lens for relating design choices (e.g., immersion and interactivity) to learners' experiences and outcomes.

\subsection{Impact of Immersion on Learning Experiences}
\label{subsec:immersion-impact}

\begin{table}[t]
	\caption{Papers that explicitly quote a learning theory or framework, and those that do not. Note that some papers appear in more than one category, as they refer to multiple theories.}
	\label{tab:learning-theories-papers}
	\centering
	\renewcommand{\arraystretch}{1.5}
	\begin{tabularx}{0.9\columnwidth}{YY}
		\toprule
		\textbf{Learning Theory} & \textbf{References}  \\
		\midrule
		\textbf{Cognitive Theory of Multimedia Learning (CTML)} 
		& \citep{parong2020, pollard2020, dewitte2024, thompson2021, johnson2022, araiza-alba2021, tang2022}  \\
		\textbf{Constructivism} 
		& \citep{checa2023, madden2020, johnson-glenberg2021}  \\
		\textbf{Embodied Cognition} 
		& \citep{madden2020, araiza-alba2021, johnson-glenberg2021}  \\
		\textbf{Cognitive Affective Model of Immersive Learning (CAMIL)} 
		& \citep{petersen2022, kaplan-rakowski2024, mulders2023}  \\
		\textbf{No Theory} 
		& \citep{cecil2018, maldonado2016, preukschas2024, smith2018, makransky2019b, rai2019, makransky2018, klingenberg2020, kozhevnikov2013, parmar2023, oberdorfer2019, loup-escande2017, murcia-lopez2016, krokos2019, hejtmanek2020, zhao2020, ochs2022, raya2018, johnson2020}  \\
		\bottomrule
	\end{tabularx}
\end{table}

Building on immersion and presence, we review comparative studies (typically IVR vs. DVR) on how immersion shapes learning. To make topical patterns explicit, we group prior work into six domains: (1) medical/health training~\citep{cecil2018, maldonado2016, preukschas2024, smith2018, petersen2022}; (2) science/biology learning~\citep{makransky2019b, madden2020, rai2019, makransky2018, johnson-glenberg2021, klingenberg2020}; (3) technical, computational, and problem-solving skills~\citep{kozhevnikov2013, checa2023, parmar2023, johnson2022, araiza-alba2021, oberdorfer2019, loup-escande2017}; (4) spatial learning~\citep{murcia-lopez2016, krokos2019, parong2020, pollard2020, zhao2020, hejtmanek2020}; (5) history/geography~\citep{dewitte2024, mulders2023, ochs2022, raya2018}; and (6) arts/music~\citep{johnson2020, tang2022, kaplan-rakowski2024}. Across domains, results are mixed: IVR often increases presence/engagement, but does not consistently improve (and can sometimes impair) learning relative to DVR.

\emph{Domain patterns.} \emph{Medical and health} uses IVR to safely simulate high-fidelity, risky scenarios~\cite{radianti2020}. Several studies report comparable knowledge gains in IVR and DVR, with both outperforming no-training controls~\citep{cecil2018, smith2018, preukschas2024}; e.g.,~\citet{cecil2018} compared an orthopedic simulator using a Geomagic Touch\footnote{\emph{https://it.3dsystems.com/haptics-devices/touch}.} haptic controller against an IVR condition and found no difference between trained groups. By contrast,~\citet{maldonado2016} reported IVR advantages for recognizing eating-disorder symptoms. In a CAMIL-based analysis,~\citet{petersen2022} found immersion increased situational interest but not knowledge about viral diseases, illustrating a recurring dissociation between affect/motivation and test performance.
\emph{Science and biology} is similarly inconsistent: DVR outperformed IVR in biology simulations in~\citet{makransky2019b} and~\citet{rai2019}, with EEG evidence of higher cognitive load in IVR~\cite{makransky2019b} (potentially amplified by language demands, as the simulation was in English for many non-native speakers). Other work on comparable content reports IVR advantages~\citep{makransky2018, johnson-glenberg2021, klingenberg2020}. In astronomy (lunar phases),~\citet{madden2020} found no differences among IVR, DVR, and a hands-on control condition.
\emph{Technical/computational/problem solving} shows benefits when interaction and embodiment align with the target skill: IVR improved understanding of relative motion~\cite{kozhevnikov2013} and abstract reasoning in the River Crossing Game (a dynamic-programming puzzle involving constrained transport and conflict avoidance)~\cite{araiza-alba2021}. In computer assembly training, both interactive conditions (IVR, DVR) outperformed passive observation but did not differ from each other~\cite{checa2023}, consistent with CAMIL's emphasis on interaction. Embodiment can be decisive: full-body-tracked IVR (with avatar) outperformed non-embodied IVR and DVR in computational thinking~\cite{parmar2023}. Interaction modality also moderates outcomes:~\citet{johnson2022} found no overall immersion effect, but learners with low spatial ability benefited most from gesture-based IVR. Other studies report no IVR--DVR differences~\citep{oberdorfer2019, loup-escande2017}.
\emph{Spatial learning} often favors IVR due to egocentric exploration, with gains in spatial memory/navigation reported in multiple studies~\citep{murcia-lopez2016, krokos2019, parong2020, pollard2020}; however, null results appear when locomotion techniques (e.g., teleportation vs. continuous movement) and task specifics limit potential advantages~\citep{zhao2020, hejtmanek2020}, suggesting sensitivity to movement affordances and assessment alignment.
\emph{History and geography} frequently show higher presence, empathy, and appreciation in IVR without reliable recall gains:~\citet{mulders2023} reported higher presence/evaluation in IVR for learning about Anne Frank's life, but lower factual learning than DVR, similarly with~\citet{ochs2022} and~\citet{raya2018}'s work. With Google Earth-based activities, IVR and DVR can perform similarly on geography content~\cite{dewitte2024}.
\emph{Arts and music} indicates that immersion helps most when gestural mapping and immediate perceptual feedback are central: IVR improved theremin training relative to DVR and traditional practice without visual aids~\cite{johnson2020}. By contrast, art-history learning showed no IVR advantage~\cite{tang2022}, and integrated audio in DVR produced the strongest outcomes in~\citet{kaplan-rakowski2024}, emphasizing multisensory design rather than immersion per se.

\emph{Synthesis and implications.} Overall, immersion effects on learning are ambiguous: IVR reliably increases presence/interest, but learning benefits depend on cognitive-load management (e.g., language demands, interface complexity)~\cite{makransky2019b}, embodiment and interaction modality (e.g., full-body avatars; gesture vs. voice)~\citep{parmar2023, johnson2022}, task--technology fit (e.g., when spatial exploration is instrumental to the learning goal)~\citep{murcia-lopez2016, krokos2019}, and the coherence of system contingencies with learners' expectations. Practically, these findings argue for match rather than maximal immersion: align locomotion and input schemes with target skills, scaffold to reduce extraneous load, and use embodiment when it directly serves the learning objective.

A final trend concerns theoretical grounding. Consistent with prior reviews~\cite{radianti2020}, many studies do not articulate an explicit learning theory (\autoref{tab:learning-theories-papers}). When theory is used, CTML supports load-sensitive multimedia design~\citep{parong2020, pollard2020, dewitte2024}; constructivism and embodied cognition motivate active, situated engagement~\citep{checa2023, madden2020, araiza-alba2021, johnson-glenberg2021}; and CAMIL frames how immersion interacts with motivational and cognitive processes~\citep{petersen2022, kaplan-rakowski2024, mulders2023}. Stronger theory--method coupling is needed to explain mixed results and specify when and how immersion should be introduced. 
In sum, immersion does not always benefit learning. In fact, the most robust gains arise when immersive affordances directly support the targeted knowledge or skill, are delivered through coherent, low-load contingencies, and are grounded in an explicit theoretical rationale.

The published studies typically examined single applications and, although qualitatively comparable, cannot be quantitatively compared due to differing protocols. Here, we use the CAMIL framework and a common experimental protocol to investigate immersion and learning across three domains: cultural heritage, environmental awareness, and high school physics. 
\section{Methods}
\label{sec:methods}

\subsection{Overview and Rationale}
\label{subsec:method-overview}

We adopted a horizontal, cross-domain methodology to examine how immersion shapes learning across three learning domains -- cultural heritage (\emph{RHOME-VR}; \autoref{sec:rhome-vr}), environmental awareness (\emph{Envisioning Corals}; \autoref{sec:ec}), and physics (\emph{Physics Playground}; \autoref{sec:pp}). Prior publications have presented the core systems and preliminary partial results for the three projects~\citep{clerici2022, boffi2023domus, boffi2023ec, battipede2025, giangualano2025}. In this section, we introduce common methodological elements; project-specific details are presented in separate sections for every learning topic.

Each study uses the same evaluation backbone -- learning, user experience, technology acceptance, sense of presence/embodiment, and cybersickness -- supplemented with domain-specific tasks and measures. The same content is delivered in all applications at different immersion levels (IVR vs. DVR), with control conditions when applicable. 
This cross-study consistency enables synthesis while respecting domain constraints (e.g., body ownership/embodiment for \emph{Envisioning Corals}; factual learning acquisition for \emph{Physics Playground} and \emph{RHOME-VR}). All applications target upper-secondary and early-tertiary learners; \autoref{tab:apps-summary} summarizes domains, experimental conditions, and core measures.

\begin{table*}[t]
	\caption{Applications, experimental conditions, and studies carried out.}
	\label{tab:apps-summary}
	
	\begin{footnotesize}
		\centering
\begin{tabularx}{\textwidth}{lll}
			\toprule
			\textbf{Application (Domain)} & \textbf{Conditions} & \textbf{Studies} \\
			\midrule
\addlinespace
			
			\parbox{.2\textwidth}{\emph{RHOME-VR}\\(Cultural heritage)} &
			\parbox{.2\textwidth}{IVR, DVR, Slideshow} & 
			\parbox{.5\textwidth}{Hybrid, between-subjects: IVR (lab) vs. DVR (remote) vs. slideshow (remote).} \\
\addlinespace
			\parbox{.2\textwidth}{\emph{Envisioning Corals}\\(Environmental awareness)} & 
			\parbox{.2\textwidth}{IVR (embodiment);\\IVR \& DVR (learning)} & 
			\parbox{.5\textwidth}{S1: IVR-only, within-subjects (avatar embodiment + EDA).\\
				S2: Between-subjects IVR vs. DVR (learning + logs + EDA).} \\
\addlinespace
			\parbox{.2\textwidth}{\emph{Physics Playground}\\(Physics education)} & 
			\parbox{.2\textwidth}{IVR \& Slideshow (lab);\\IVR, DVR, control (ecological)} &
			\parbox{.5\textwidth}{S1: Lab mixed-methods (IVR vs. slideshow + expert/teacher input).\\
				S2: High school pre--post--retention (IVR vs. DVR vs. 2D).}\\			
			\bottomrule
		\end{tabularx}
	\end{footnotesize}	
\end{table*}

\subsection{Shared Methodology}
\label{subsec:methodology}

Across studies, we kept common methodological elements. Primary recruitment targeted young adults (18--35), with instructional content and audio language matched to participants' native language to minimize extraneous load. Studies were conducted in laboratory settings to maximize experimental control, and study protocols were evaluated and approved by a university ethics committee before the experiment began. When movement was required, locomotion used teleportation to mitigate cybersickness~\cite{clifton2020}. 

The evaluation backbone was consistent across applications: learning was assessed with custom multiple-choice tests aligned to topic objectives (Roman architecture; coral-reef ecology and pro-environmental strategies; kinematics/gravity); technology acceptance and user experience were measured via the Technology Acceptance Model (TAM)~\citep{davis1989, rese2017, salloum2019} and either AttrakDiff Pragmatic/Hedonic Quality~\citep{hassenzahl2004, sagnier2020} or the User Experience Questionnaire (UEQ, either the full~\cite{ueq2008} or the short~\cite{ueq-s2017} version); presence and embodiment were captured with Witmer and Singer's Presence Questionnaire~\cite{witmer1998, witmer2005} and the Avatar Embodiment Questionnaire~\cite{gonzalez-franco2018}; cybersickness was measured with the Virtual Reality Sickness Questionnaire (VRSQ;~\cite{kim2018}); workload with NASA Task Load indeX (TLX~\cite{tlx-nasa}) and usability with the System Usability Scale (SUS~\cite{sus}) where relevant.

Sessions followed a shared protocol: informed consent and demographics/experience, the assigned condition(s) with a short in-app tutorial, and post-exposure questionnaires in a fixed order. Where applicable, logs and physiological signals were collected during application use.

\section{RHOME-VR}
\label{sec:rhome-vr}

\subsection{Application's Design}
\label{subsec:rh_design}

\begin{figure}[t]
	\centering
	\includegraphics[width=\linewidth]{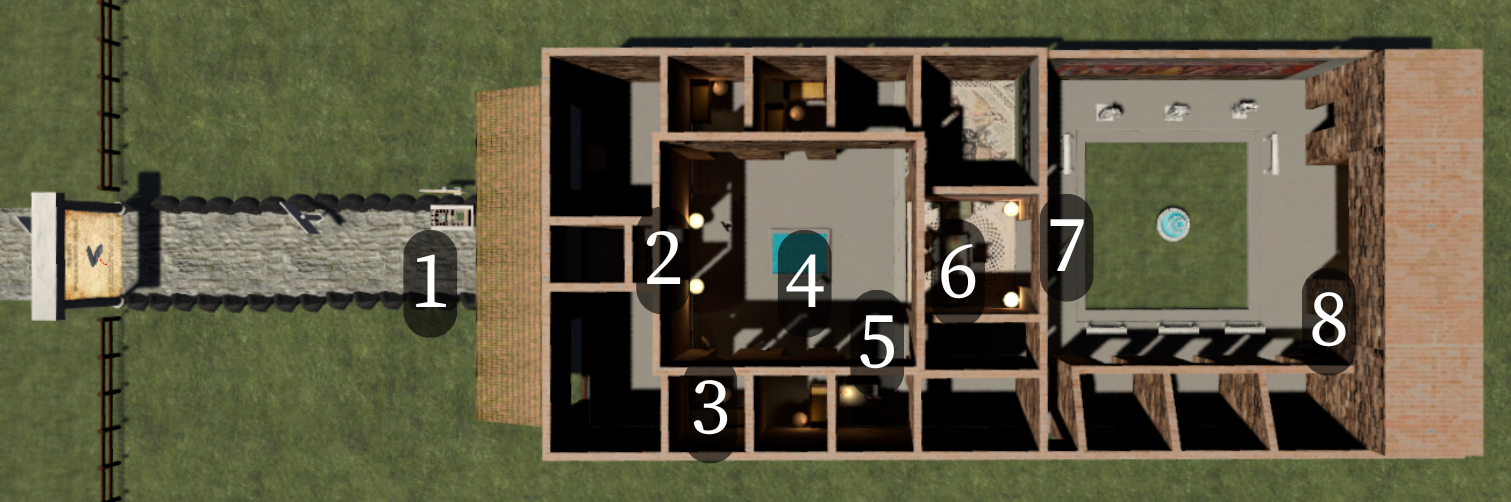}
	\caption{\emph{RHOME-VR} map with eight lecture points.}
	\label{fig:rhome-map}
\end{figure}

\emph{RHOME-VR} is a virtual tour of a Roman domus presented and evaluated by Clerici et al. and Boffi et al.~\citep{boffi2023domus, clerici2022}. It allows users to explore a reconstructed Roman Domus through a guided, eight-step tour mimicking virtual museums. Learners advance along a fixed path (\autoref{fig:rhome-map}); at each `lecture point' a green marker indicates the target spot, alongside a signboard naming each room. Triggering the instructional audio starts a voiced narration while movement and signboards are temporarily disabled to focus attention. 

\subsection{Experimental Design}
\label{subsec:rh_exp}

A between-subjects design study was used to evaluate the effectiveness of this approach in transmitting historical knowledge, with the media as a between-subjects factor. Participants were randomized into one of three conditions: Immersive Virtual Reality (IVR), Desktop Virtual Reality (DVR), and control. The dependent variables were learning outcome (factual knowledge and spatial learning), user experience, technology acceptance, presence, and cybersickness -- the last two measures we collected only for the VR experiences. 

The data collection took place during the COVID-19 pandemic. A hybrid testing setup was adopted to avoid placing participants at risk while still achieving a statistically reliable sample size: IVR participants were tested on site in the laboratory, whereas DVR and control conditions were delivered remotely. This arrangement, however, provided less control over the remote conditions.

The dependent variables were measured through self-reported questionnaire-based measures. The survey was based on similar literature case studies~\citep{kyrlitsias2020, sagnier2020}. The learning part of the questionnaire -- developed in collaboration with a former high-school history teacher -- was composed of factual and spatial knowledge parts. Factual knowledge consisted of multiple-choice history questions, while the spatial knowledge was built over two different tasks. In the first, users had to assign the names of the rooms on an empty Domus map (\emph{Rooms Learning}), while in the second, users should properly link objects to the room in which they appeared (\emph{Objects Learning}). The other dependent variables were measured through validated literature constructs, collecting technological acceptance, user experience, presence, and cybersickness data. 
The control condition consisted of a slideshow mirroring the VR application's structure using screenshots from the VE and the same instructional audio. 

\begin{figure*}[t]
	\captionsetup[sub]{font=scriptsize}
	\begin{footnotesize}
		\centering
\begin{subfigure}{0.16\linewidth}
			\centering
			\includegraphics[width=\linewidth]{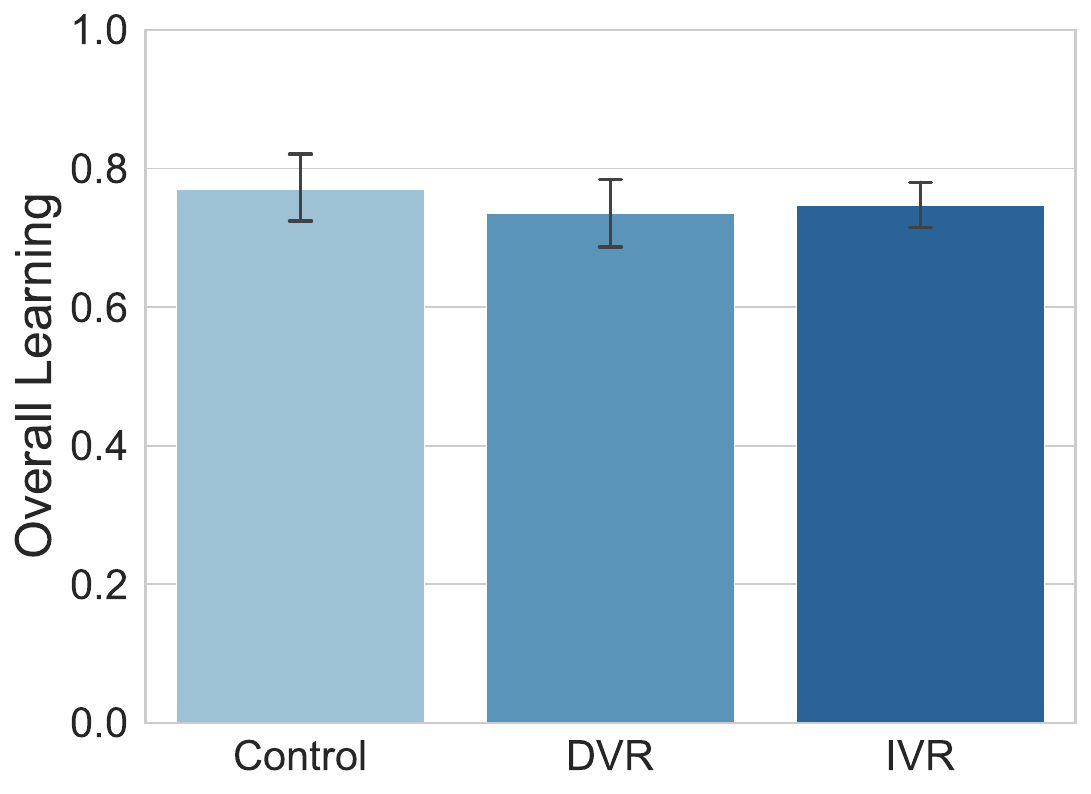}
			\subcaption{Overall learning.}
			\label{subfig:rhome-general-learning}
		\end{subfigure}
\begin{subfigure}{0.16\linewidth}
			\centering
			\includegraphics[width=\linewidth]{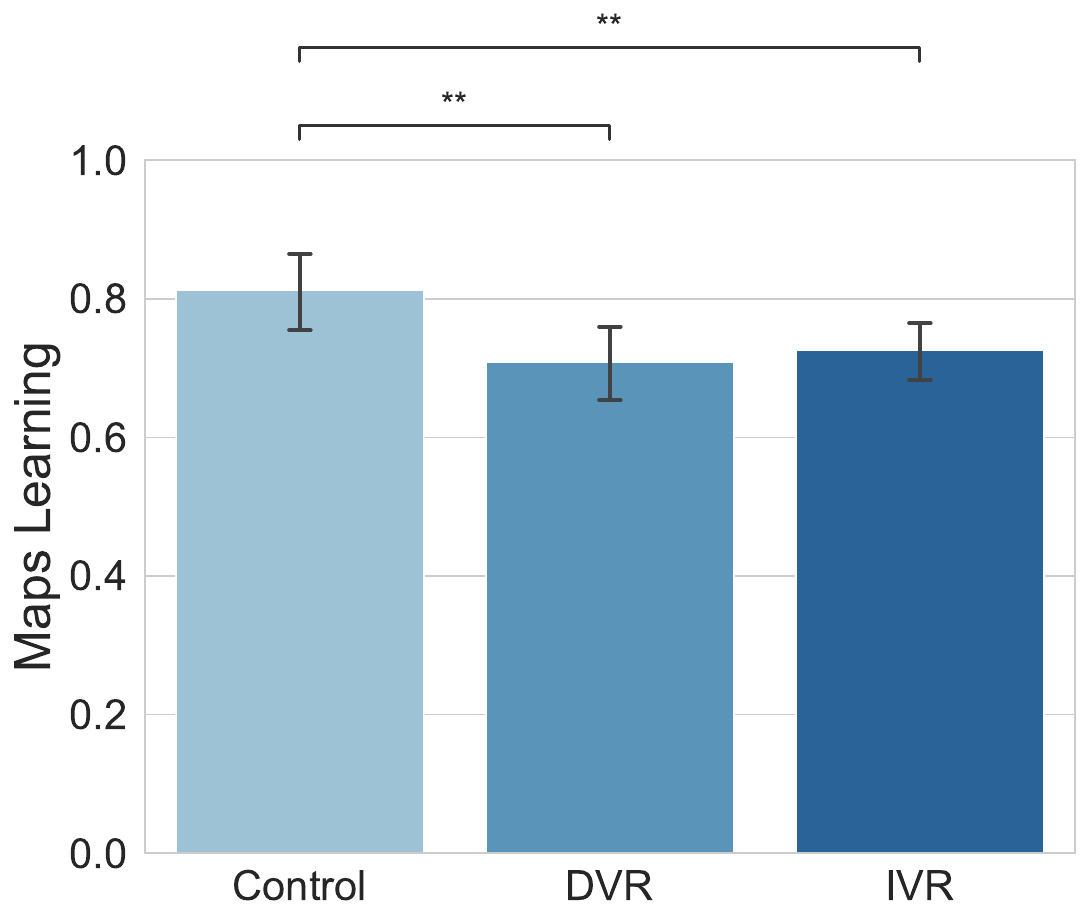}
			\subcaption{Maps learning.}
			\label{subfig:rhome-maps-learning}
		\end{subfigure}
\begin{subfigure}{0.16\linewidth}
			\centering
			\includegraphics[width=\linewidth]{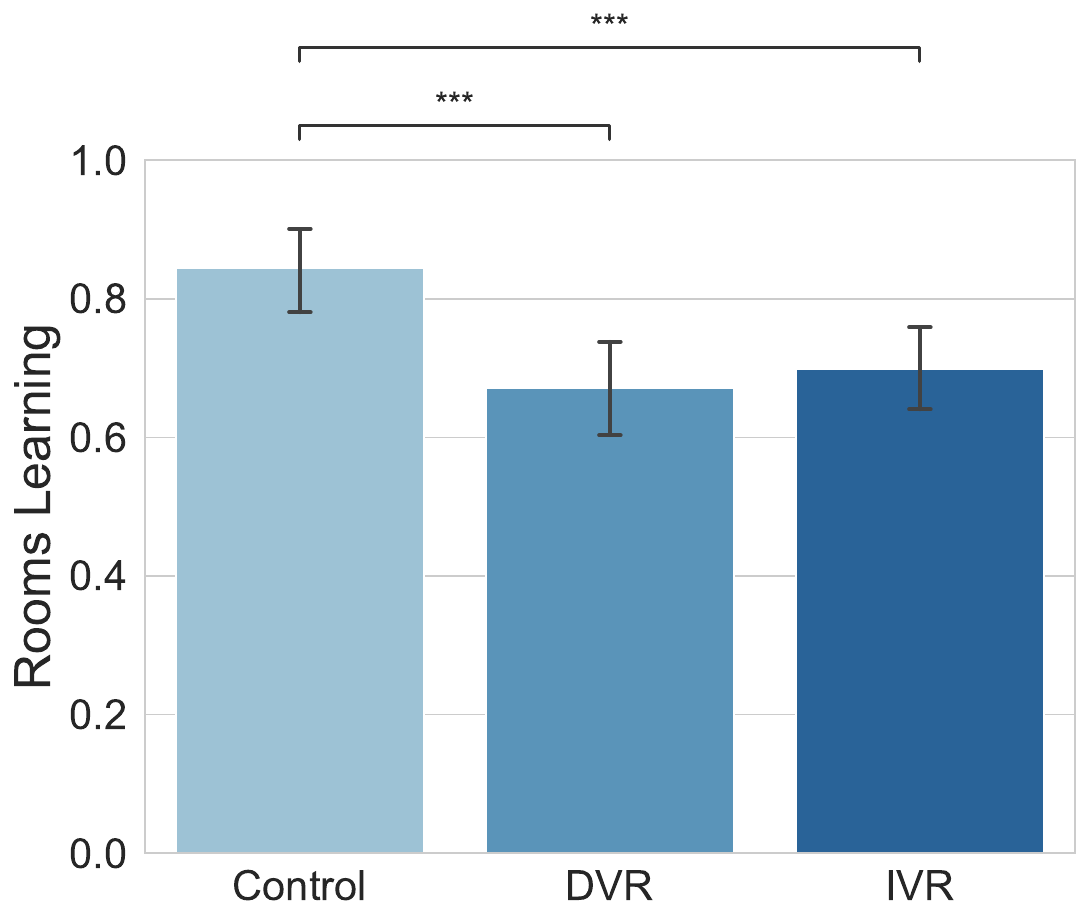}
			\subcaption{Rooms learning.}
			\label{subfig:rhome-rooms-learning}
		\end{subfigure}
\begin{subfigure}{0.16\linewidth}
			\centering
			\includegraphics[width=\linewidth]{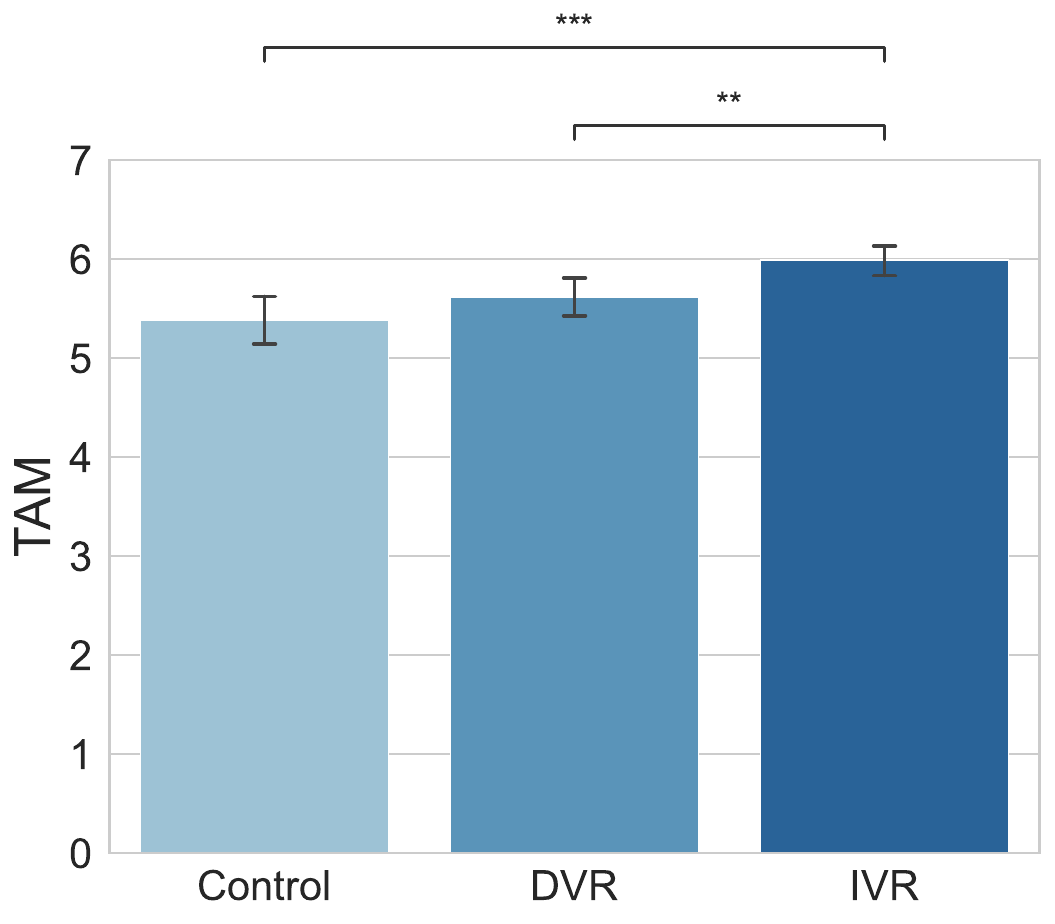}
			\subcaption{Overall TAM.}
			\label{subfig:rhome-tam}
		\end{subfigure}
\begin{subfigure}{0.16\linewidth}
			\centering
			\includegraphics[width=\linewidth]{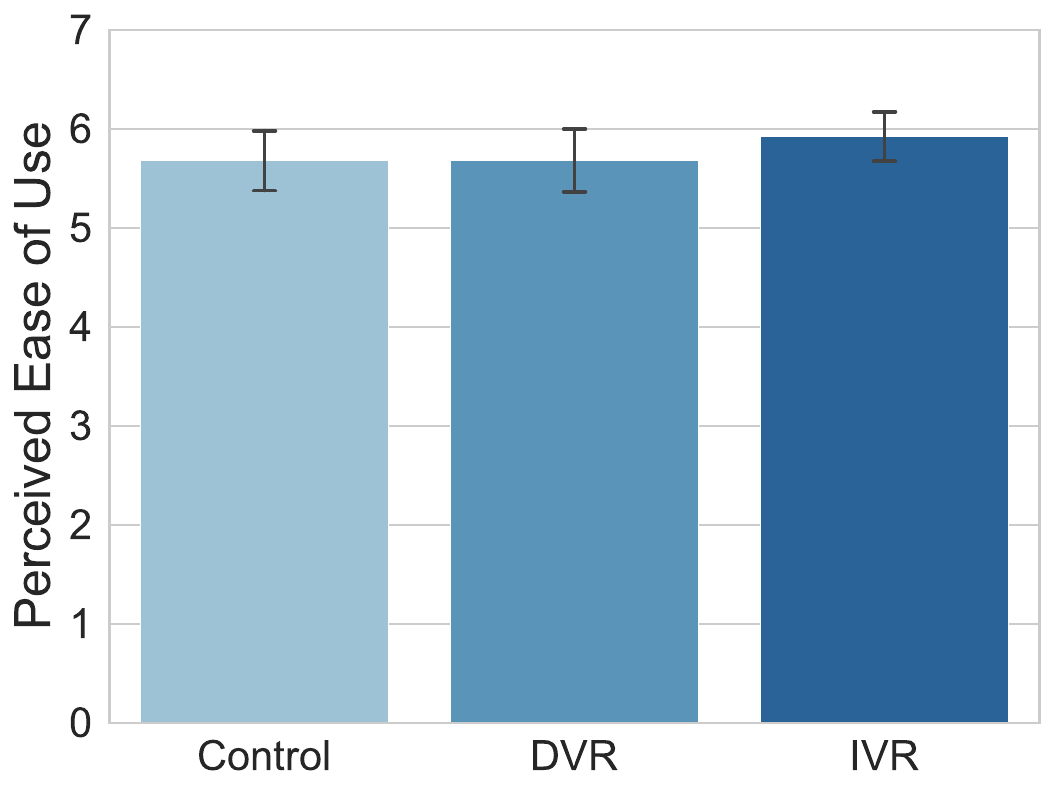}
			\subcaption{Perceived Ease of Use.}
			\label{subfig:rhome-peu}
		\end{subfigure}
\begin{subfigure}{0.16\linewidth}
			\centering
			\includegraphics[width=\linewidth]{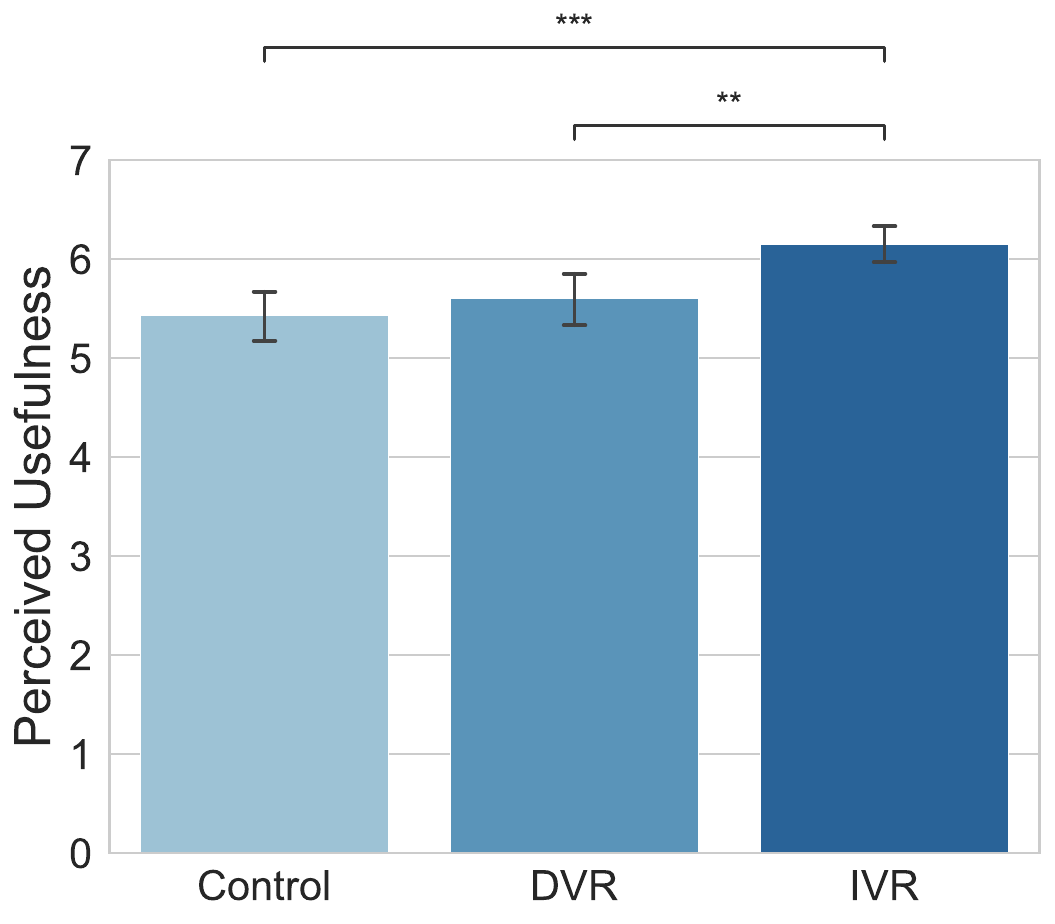}
			\subcaption{Perceived Usefulness.}
			\label{subfig:rhome-pu}
		\end{subfigure}
	\end{footnotesize}
	
\caption{\emph{RHOME-VR} learning ((a), (b), and (c)) and TAM ((d), (e), (f)) results.}
	\label{fig:rhome-learning}
	\label{fig:rhome-tam}
\end{figure*} 
\subsection{Results}
\label{subsec:rh_results}
161 individuals participated in the study (89 female, 72 male), all aged between 18 and 35 ($M = 23.8$, $sd = 2.9$). 
Overall learning did not differ by condition (\autoref{subfig:rhome-general-learning}). 
Mann-Whitney tests against control were not significant (IVR vs control: $U = 1254.5$, $p = .16$; DVR vs control: $U = 1199$, $p = .25$), as well as a Welch's Anova between IVR and DVR ($F(1, 106) = 0.135$, $p = .71$). Decomposing learning into subvariables revealed significant differences between conditions in the \emph{Maps Learning} scores (\autoref{subfig:rhome-maps-learning}), indicating spatial knowledge acquisition: control participants outperformed both IVR ($U = 997$, $p = .003$; $M_{IVR} = 9.39$) and DVR ($U = 977.5$, $p = .009$). This difference was driven by the \emph{Rooms Learning} results (\autoref{subfig:rhome-rooms-learning}), where the same pattern occurred (IVR vs control: $U = 933.5$, $p < .001$; DVR vs control: $U = 864$, $p < .001$). All other pairwise comparisons revealed no significant differences.

On TAM, overall IVR scored significantly higher than both DVR and control (\autoref{subfig:rhome-tam}), as revealed by pairwise Welch's Anova tests (IVR vs control: $F(1, 107) = 16.52$, $p < .001$; IVR vs DVR: $F(1, 106) = 0.96$, $p = .003$). Control and DVR did not differ ($F(1, 103) = 2.06$, $p = .154$). At construct levels, interesting results occurred in the \emph{Perceived Ease of Use} (PEU) subscale, where a Mann-Whitney test revealed no differences between conditions (\autoref{subfig:rhome-peu}), and in the \emph{Perceived Usefulness} (PU) subscale, where IVR significantly outperformed both DVR and control (\autoref{subfig:rhome-pu}) (IVR vs DVR: $U = 1949.5$, $p = .002$; IVR vs control: ($U = 2161.5$, $p < .001$). 

In overall UX, Welch's test revealed that IVR significantly outperformed control and DVR (IVR vs control: $F(1, 107) = 12.04$, $p < .001$; IVR vs DVR: $F(1, 106) = 14.23$, $p < .001$), while no differences occurred between control and DVR ($F(1, 103) = 0.108$, $p = .742$). Subscales showed higher Pragmatic Qualities scores for IVR than DVR ($U = 1809.5$, $p = .03$), and higher Hedonic Qualities for IVR than both DVR ($U = 1909$, $p = .005$) and control ($U = 1948$, $p = .005$).    

Presence was higher in IVR than DVR, as revealed by a Welch's test ($F(1, 106) = 14.54$, $p < .001$), driven by Involvement ($F(1, 106) = 20.41$, $p < .001$) and Immersion ($F(1, 106) = 16.34$, $p < .001$); Sensory Fidelity and Interface Quality showed no differences. Cybersickness was low overall and unrelated to condition ($\chi^2(1) = 0.493$, $p = .482$).

\section{Envisioning Corals}
\label{sec:ec}
\subsection{Application's Design}
\label{subsec:ec_design}

\begin{figure*}[t]
	\centering
	\begin{subfigure}{0.329\textwidth}
		\includegraphics[width=\linewidth]{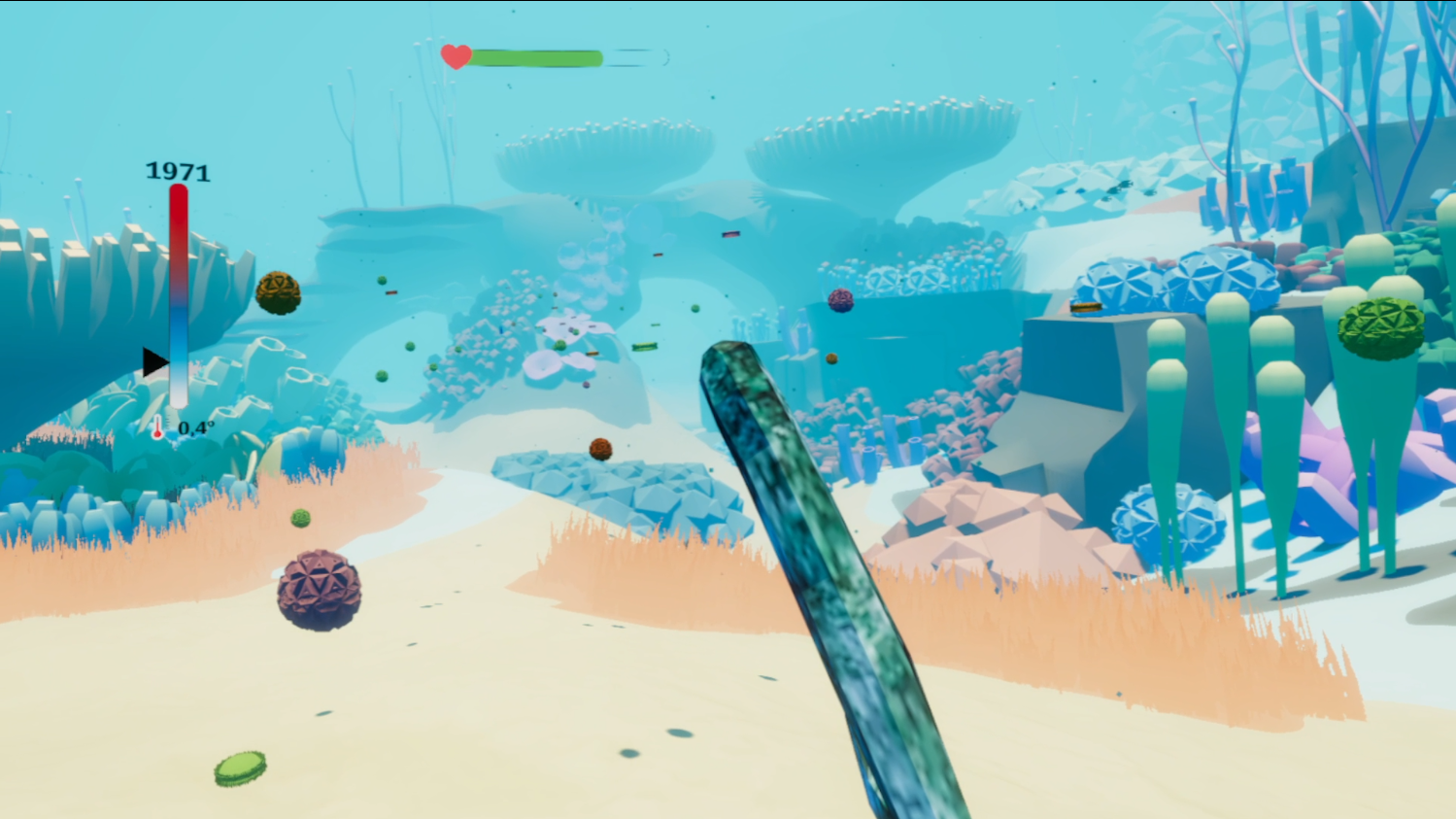}
	\end{subfigure}
\begin{subfigure}{0.329\textwidth}
		\includegraphics[width=\linewidth]{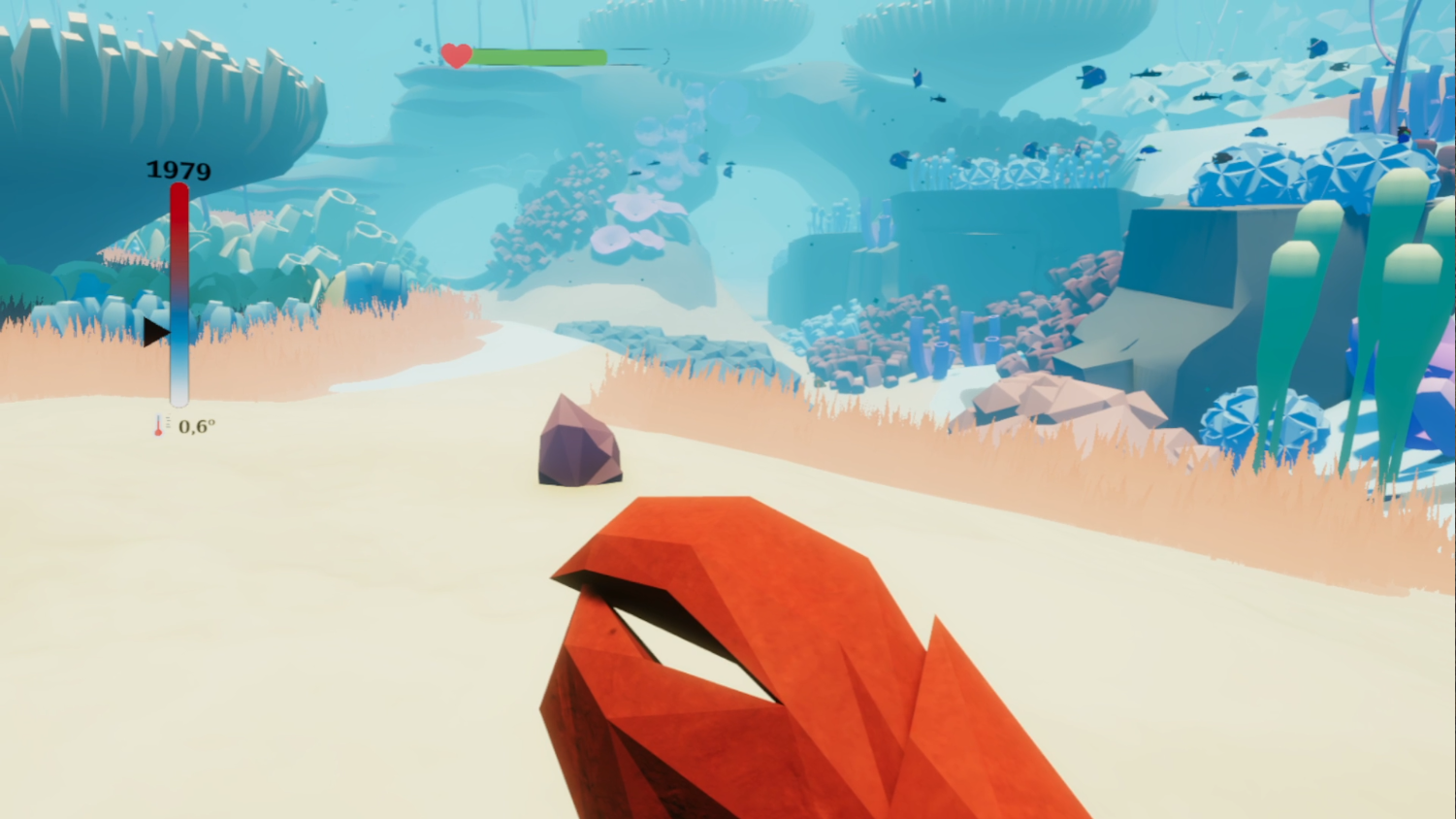}
	\end{subfigure}
\begin{subfigure}{0.329\textwidth}
		\includegraphics[width=\linewidth]{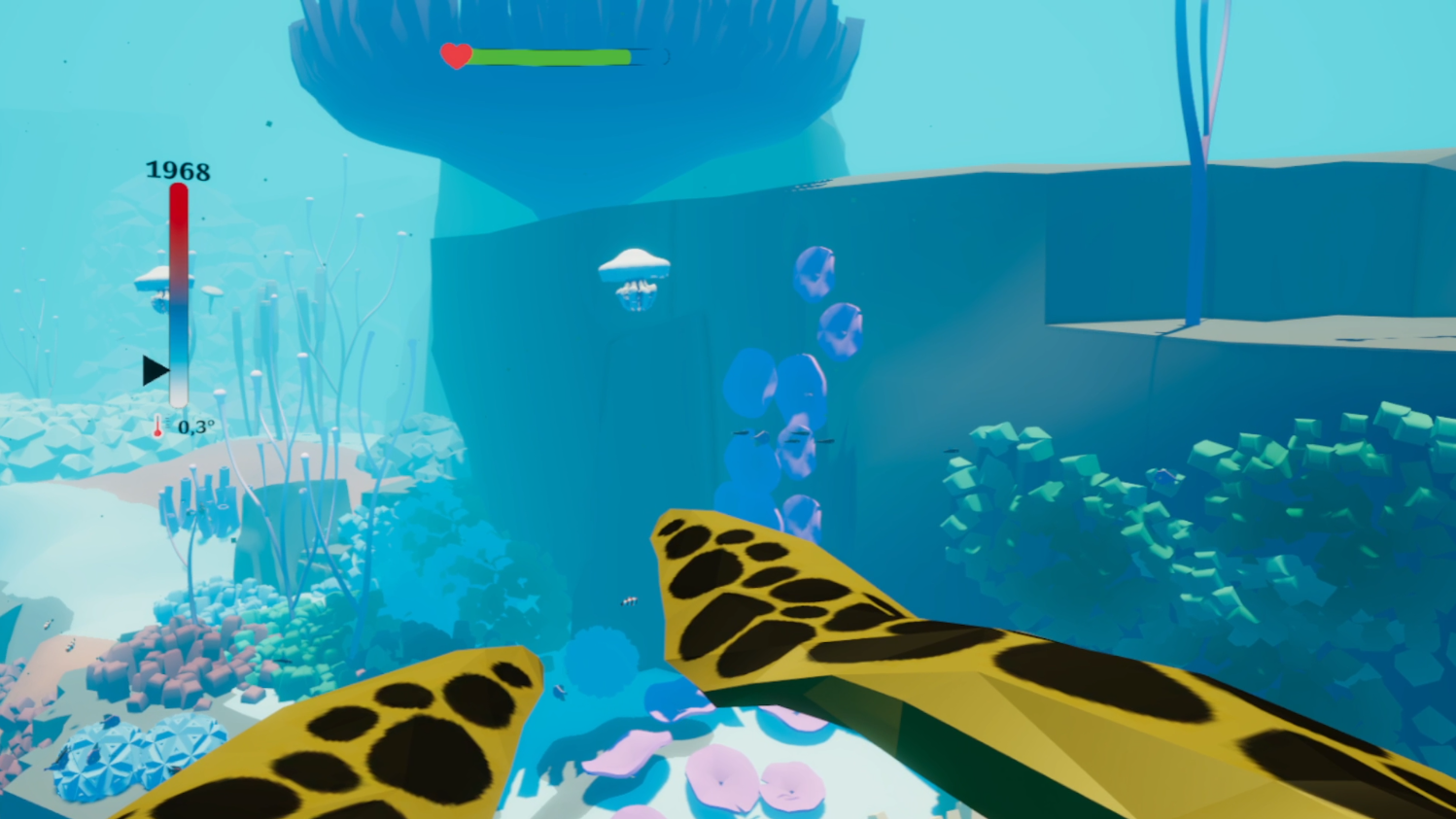}
	\end{subfigure}
\caption{Food collected by \emph{Envisioning Corals}' characters: microorganisms moved by the sea current (coral -- left), small debris (hermit crab -- middle), and jellyfish (turtle -- right).}
	\label{fig:ec-food}
\end{figure*}

\emph{Envisioning Corals} is an educational VR experience that allows users to embody one of three reef inhabitants -- a coral, a hermit crab, or a sea turtle -- while witnessing six decades of environmental changes over the coral reef~\citep{boffi2023ec, giangualano2025}. Its design follows~\citet{scurati2021}'s framework for environmental IVR (emotional, rational, practical): learners freely select an avatar (emotional bond), traverse a coral-reef scene inspired by the Maldives while a 60-year timeline advances from 1960 to the present (rational visualization of long-term effects), and receive closing guidance on actionable behaviors (practical cues), paired with a restorative scene reset to avoid a purely negative framing. Gameplay is intentionally minimal: collect species-appropriate food while plastic debris accumulates and resources become scarcer, so the rising difficulty mirrors degradation. A lightweight HUD shows a life bar, current year, and sea-temperature indicator (\autoref{fig:ec-food}); educational narration at the beginning and end frames the free-exploration window and is supported by subtitles. A dedicated tutorial area -- set in a reconstruction of a laboratory environment -- precedes the reef, with voiced guidance and unlimited practice time.

\subsection{Study 1: Embodiment Evaluation}
\label{subsec:ec_study1}

A lab study was conducted to assess how avatar characteristics modulate embodiment in IVR. The design was pure within-subjects (IVR only): each participant experienced all three avatars once, in randomized order. Self-reported measures -- an adapted version of the Avatar Embodiment Questionnaire~\cite{gonzalez-franco2018} --  were combined with physiological measures -- Electrodermal Activity (EDA). For EDA, participants were first instructed to remain still for 2 minutes, to record a baseline;  EDA was then recorded beginning 2 seconds after embodiment occurred, and continuing for a 10-second interval. 

\subsection{Study 1: Results}
\label{subsec:ec_study1-results}

\begin{figure}[t]
	\centering
\begin{subfigure}{0.35\linewidth}
		\centering
		\includegraphics[width=\linewidth]{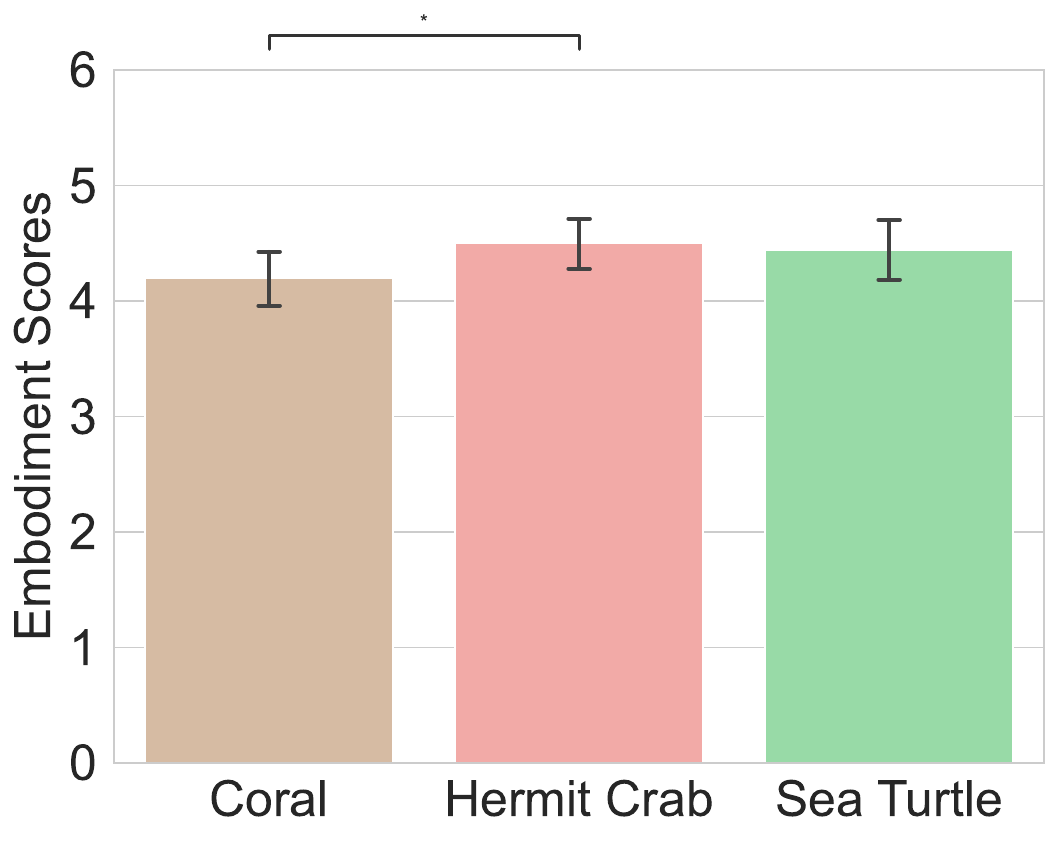}
	\end{subfigure}
\quad
\begin{subfigure}{0.35\linewidth}
		\centering
		\includegraphics[width=\linewidth]{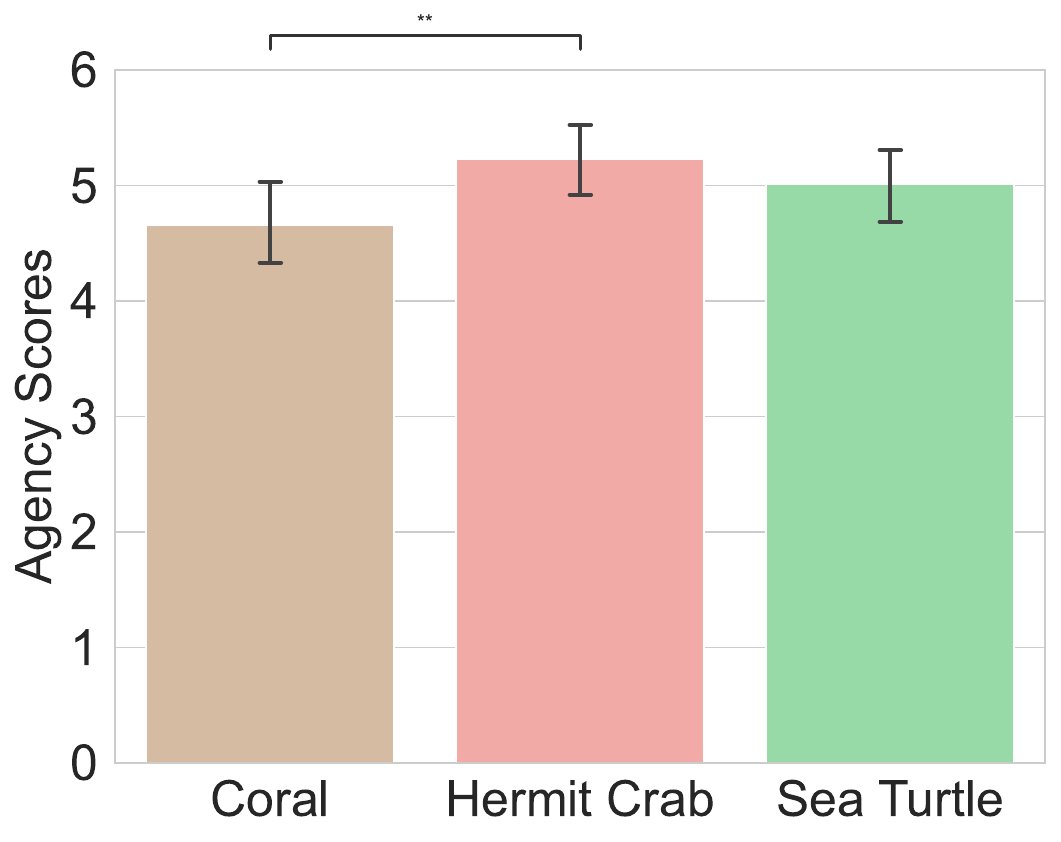}
	\end{subfigure}
\par\medskip \begin{subfigure}{0.45\linewidth}
		\centering
		\includegraphics[width=\linewidth]{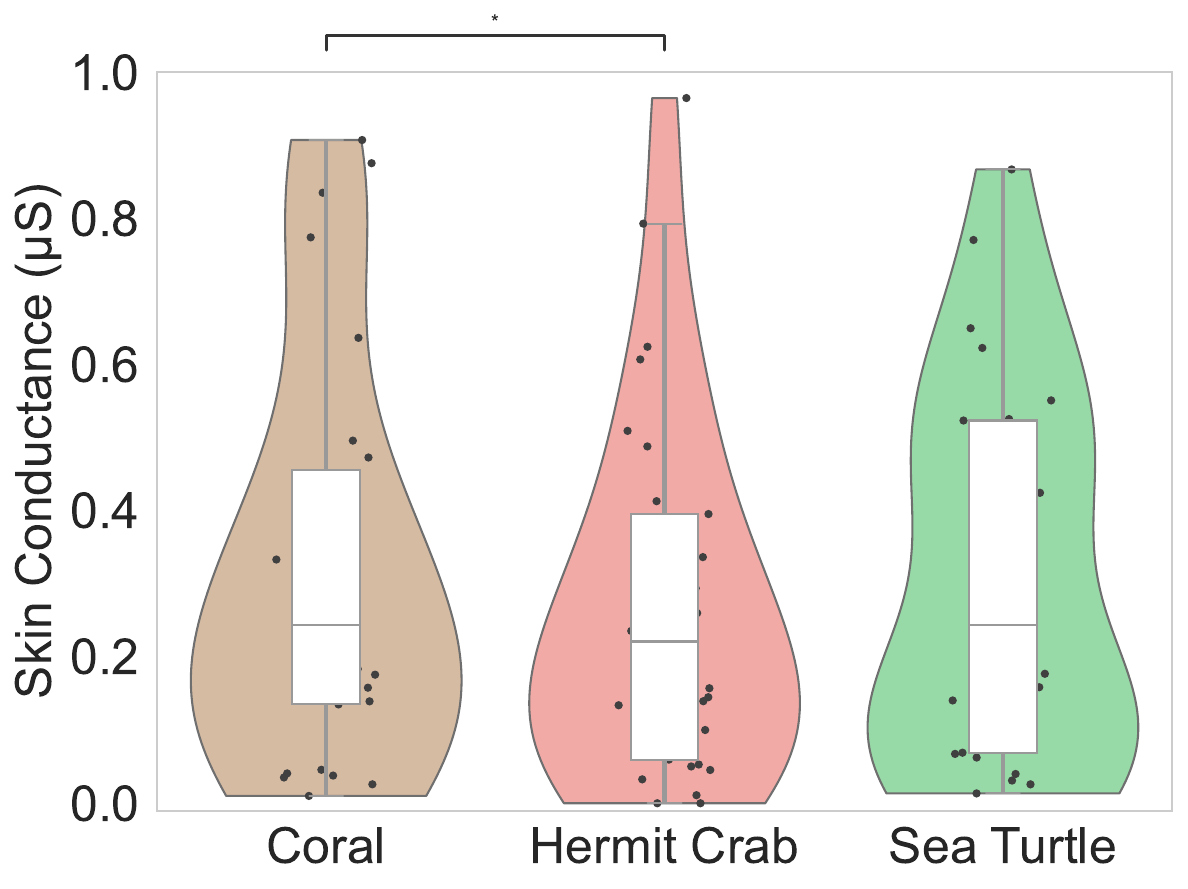}
	\end{subfigure}
\caption{\emph{Envisioning Corals} study 1 results, divided by avatar. Top: Overall embodiment and agency. Bottom: EDA.}
	\label{fig:ec_study1-embodiment-eda}
\end{figure}

$31$ participants completed the study ($18$ female, $13$ male, the majority aged between $25$ and $34$). All continuous variables were z-scaled for better statistical interpretations, and the results are displayed in \autoref{fig:ec_study1-embodiment-eda}.

A Linear Mixed Model (LMM) on overall embodiment revealed a main effect of the avatar ($\chi^2(2) = 7.47$, $p = .023$), with the hermit crab revealing significantly higher scores than the coral ($t(58.1) = -2.57$, $p = .034$); other pairs were not significant. This difference was driven by the Agency subscale ($\chi^2(2) = 10.478$, $p = .005$), with the same pattern (coral $<$ hermit crab; $t(60) = -3.207$, $p = .006$). 

EDA was then computed as the sum of significant amplitudes (i.e., higher than $0.01$ $mV$). An LMM revealed that EDA increased with higher self-reported embodiment ($\chi^2(1) = 4.433$, $p = .035$), and varied by avatar ($\chi^2(2) = 5.987$, $p = .05$). Post-hoc tests showed higher EDA for coral than hermit crab ($t(47.9) = 2.501$, $p = .041$); other pairwise differences were not significant. 

\subsection{Study 2: Learning Evaluation}
\label{subsec:ec_study2}

The second study compared immediate learning and experiential responses between IVR and DVR using a within-between subjects design, with time (pre vs. post) as a within factor and media (IVR vs. DVR) as a between factor. Given the previous results, only the hermit crab avatar was used. 

Participants were randomly divided between the media conditions. Measures included a 10-item factual knowledge test and the Avatar Embodiment Questionnaire. In-app behavior was logged -- specifically, distance traveled and food/junk collected -- and EDA was recorded throughout three different events: after the embodiment (Embodiment), at the end of the introductory instructional audio (Intro End), and before the last instructional audio regarding pro-environmental actions (Env-Actions). A 30-second EDA baseline was collected in a resting state before the beginning of the experience.

\subsection{Study 2: Results}
\label{subsec:ec_study2-results}

$27$ people completed the study ($20$ female, $5$ male, $2$ non-binary), aged between $19$ and $28$ ($M = 20.93$, $sd = 2.22$). 

\begin{figure*}[t]
	\centering
	
	\captionsetup[sub]{font=scriptsize}
	\begin{subfigure}{0.22\linewidth}
		\centering
		\includegraphics[width=\linewidth]{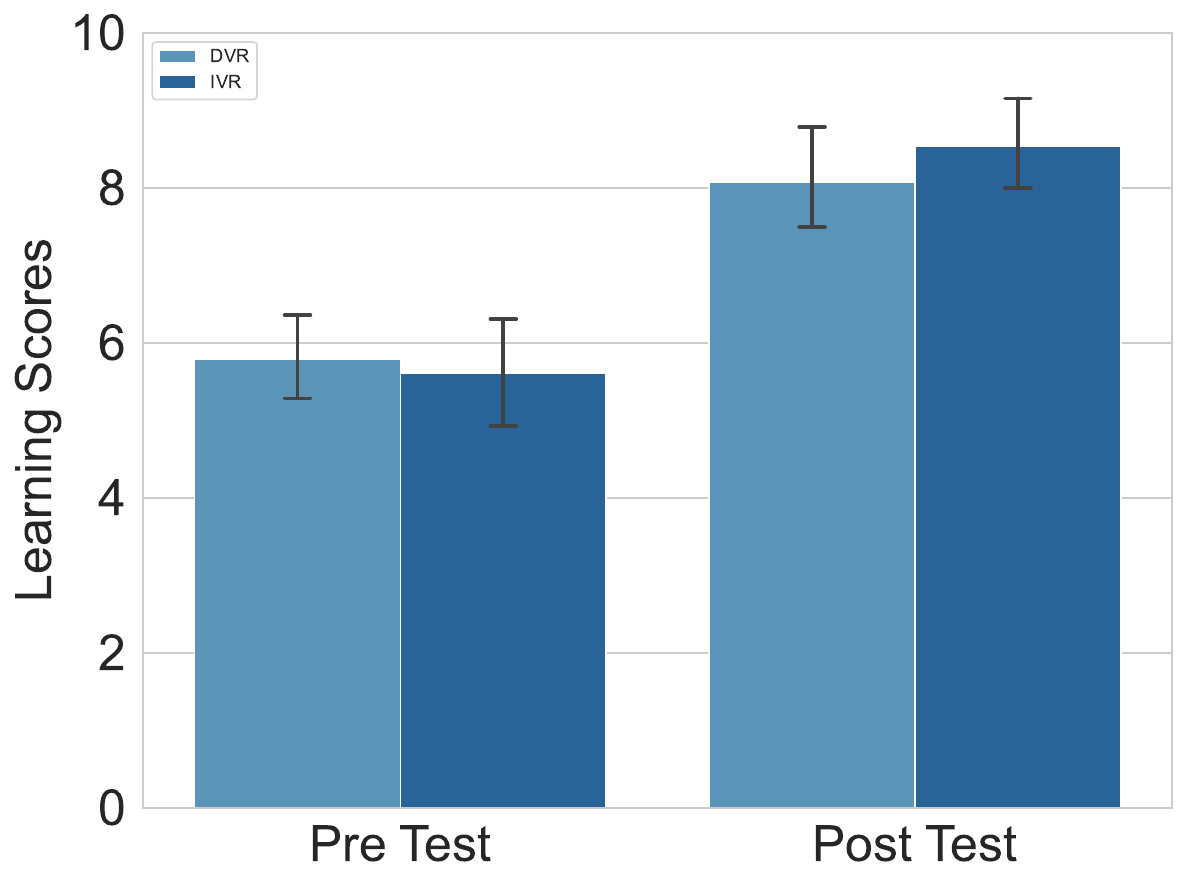}
		\subcaption{Learning}
		\label{fig:envisioning_study2_learning}		
	\end{subfigure}
\begin{subfigure}{0.22\linewidth}
		\includegraphics[width=\linewidth]{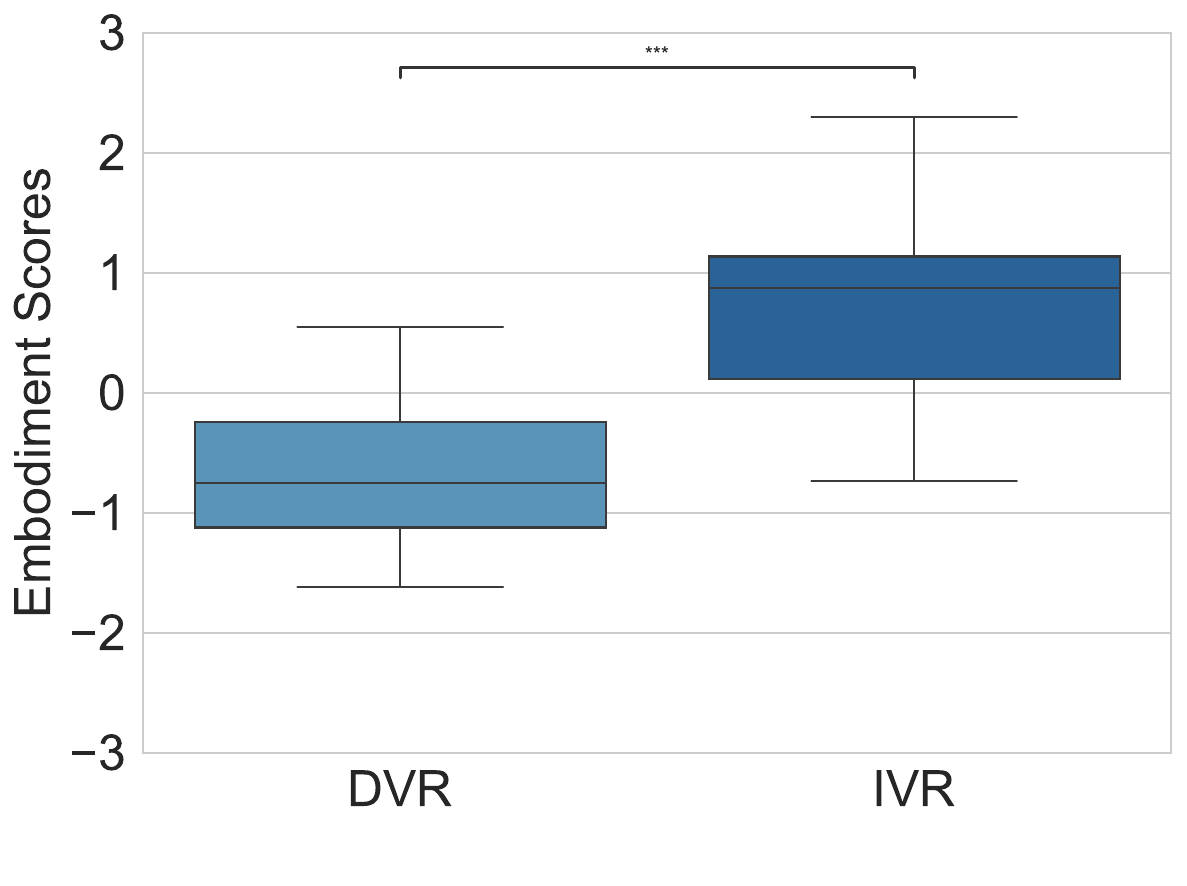}
		\subcaption{Overall Embodiment}
		\label{fig:envisioning_study2_overall_embodiment}	
	\end{subfigure}
\begin{subfigure}{0.22\linewidth}
	\includegraphics[width=\linewidth]{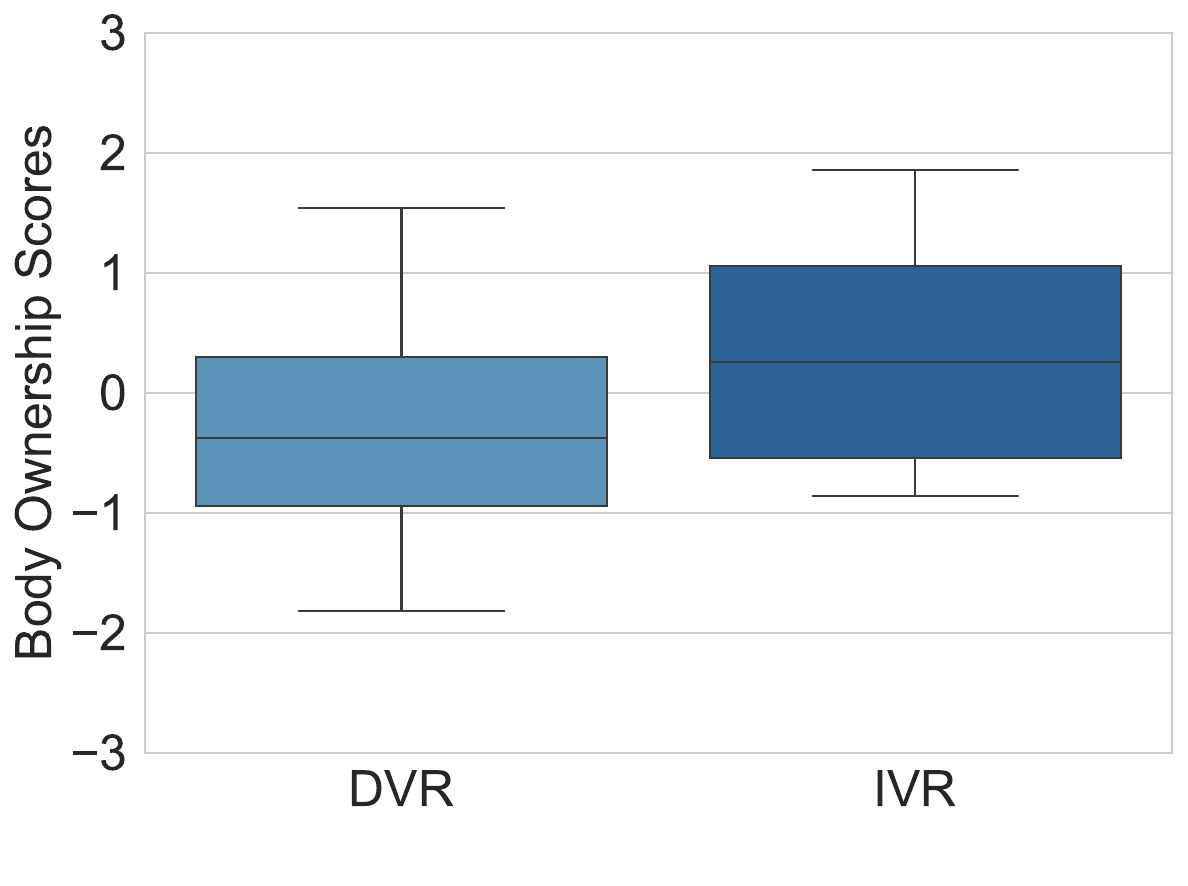}
	\subcaption{Body Ownership}
	\label{fig:envisioning_study2_body_ownership}	
	\end{subfigure}
\begin{subfigure}{0.22\linewidth}
		\includegraphics[width=\linewidth]{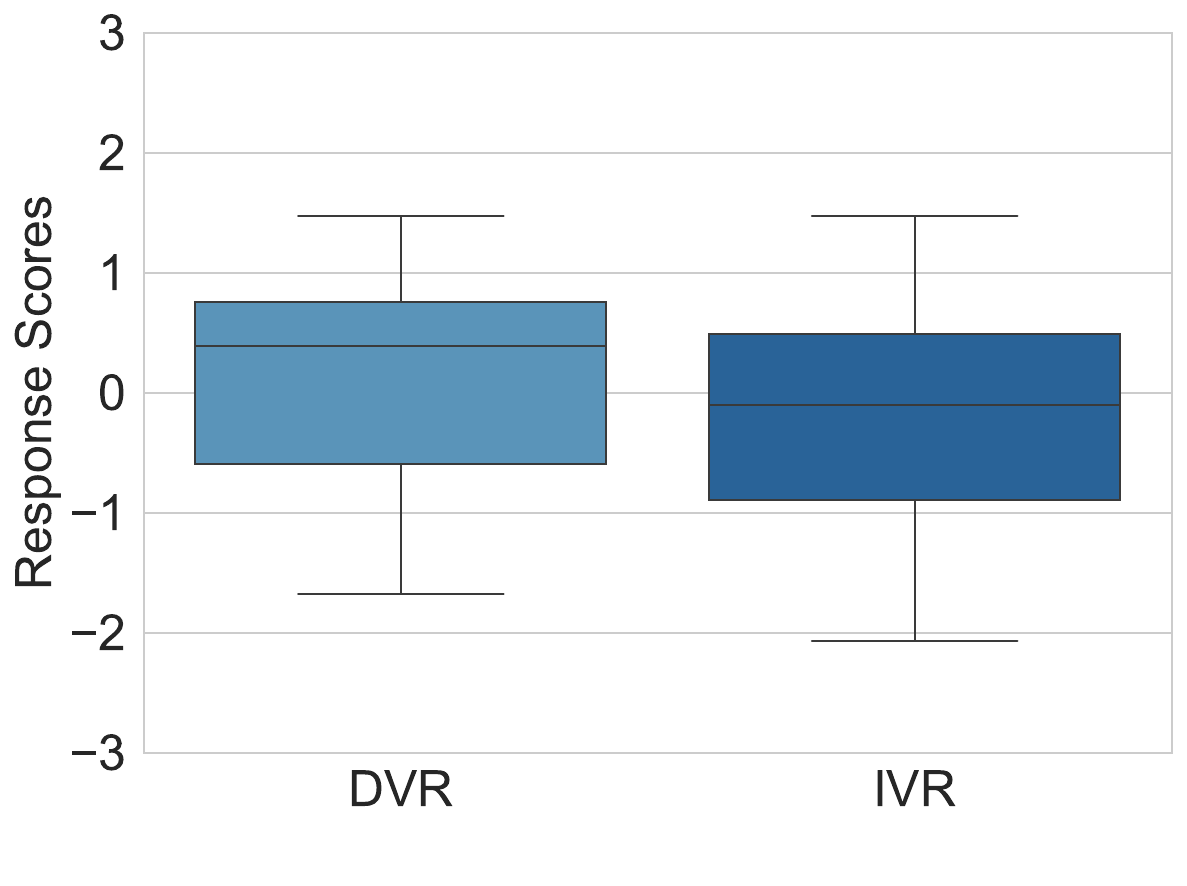}
		\subcaption{Response}
		\label{fig:envisioning_study2_response}	
	\end{subfigure}
	
	\vskip.1in	
\begin{subfigure}{0.22\linewidth}
		\includegraphics[width=\linewidth]{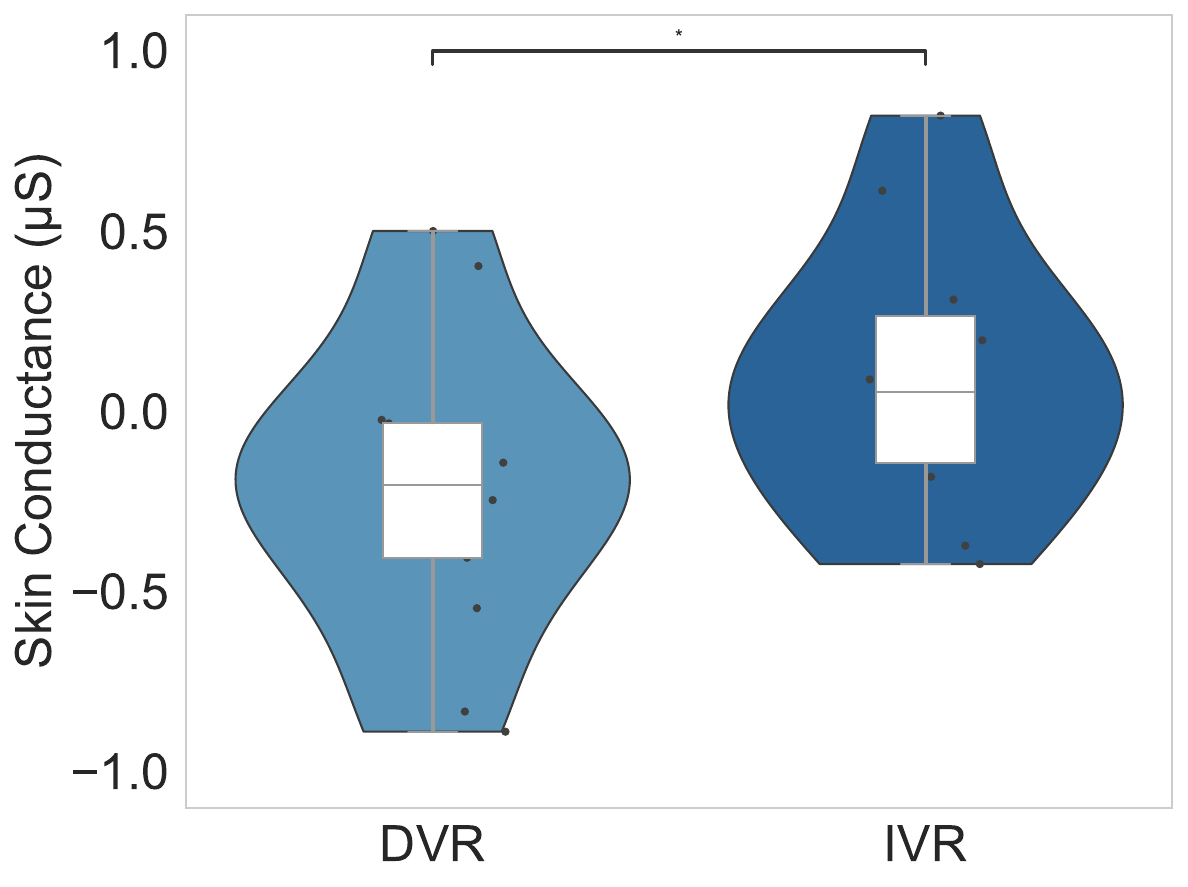}
		\subcaption{Embodiment (ESA)}
		\label{fig:envisioning_study2_eda_embodiment}
	\end{subfigure}
\begin{subfigure}{0.22\linewidth}
		\includegraphics[width=\linewidth]{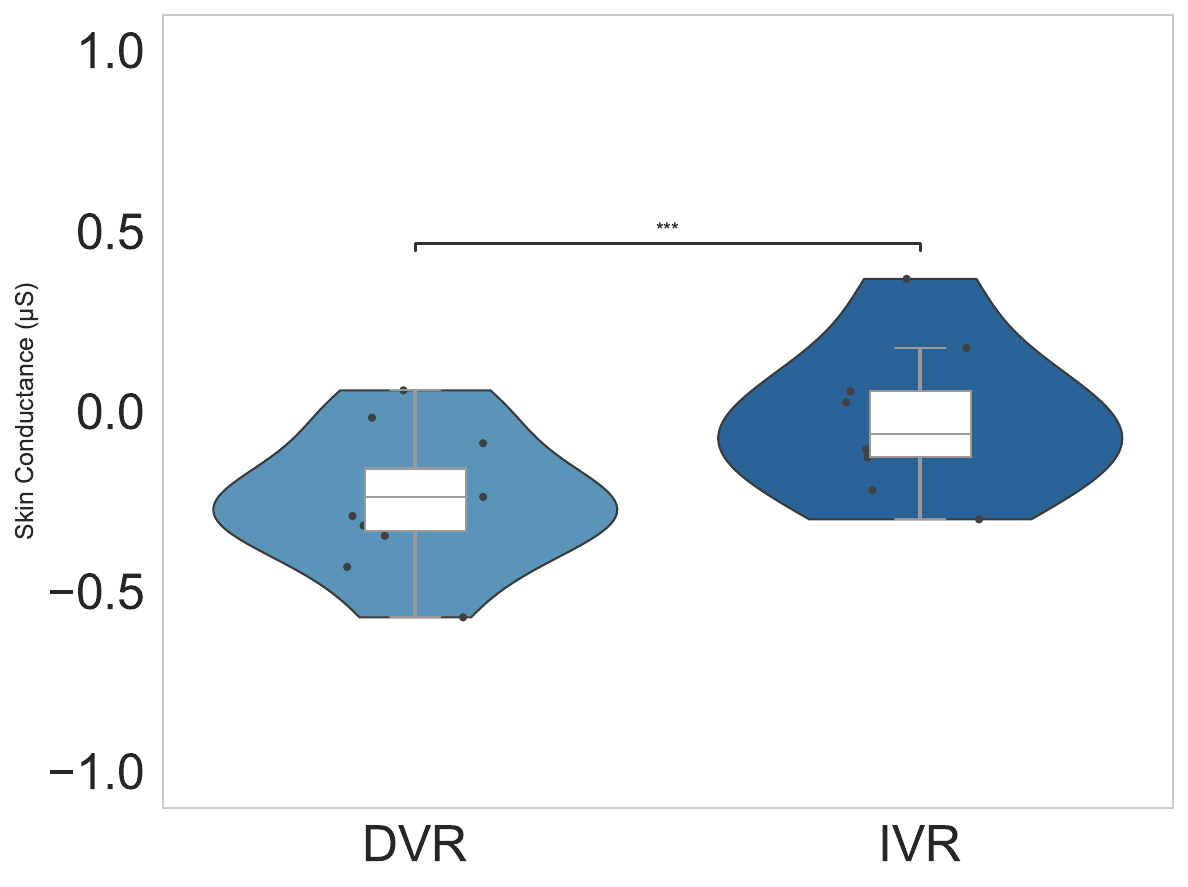}
		\subcaption{Intro-End (ESA)}
        \label{fig:envisioning_study2_eda_mov}
	\end{subfigure}
\begin{subfigure}{0.22\linewidth}
		\includegraphics[width=\linewidth]{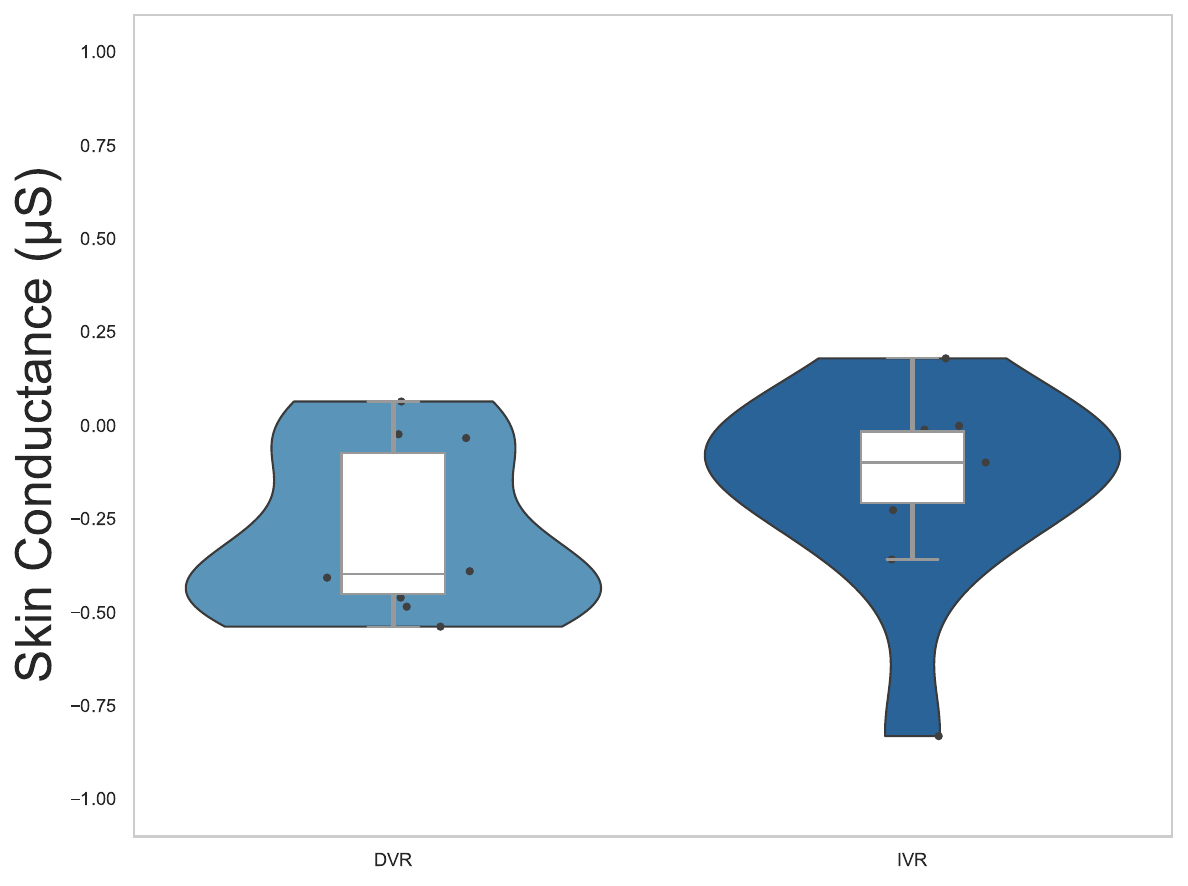}
		\subcaption{Env-Actions (ESA)}
		\label{fig:envisioning_study2_eda_end}
	\end{subfigure}
\caption{\emph{Envisioning Corals} study 2 results: (a) Learning; (b)-(g) Embodiment (Overall, Body Ownership, Response, Embodiment (EDA), Intro-End (EDA), and Env-Actions (EDA)).}
	\label{fig:ec_study2-embodiment-eda}
\end{figure*}

An LMM with the learning score as dependent variable showed a significant effect of the time point ($\chi^2(1) = 60.66$, $p < .001$), meaning that participants significantly improved their knowledge immediately after the experience (\autoref{fig:envisioning_study2_learning}). However, no effect of the media condition was found ($\chi^2(1) = 0.972$, $p = .324$), resulting in comparable learning gains across the two groups. 

Embodiment scores (Figures \ref{fig:envisioning_study2_overall_embodiment}--\ref{fig:envisioning_study2_response}) were evaluated through Student's t-test, revealing significantly higher scores for the IVR group ($t = 5.19$, $p < .001$; $M_{IVR} = 2.63$, $M_{DVR} = 1.39$). At the subscales level, we found significant differences in Agency, Tactile Sensation, Location, and Appearance subscales (all with IVR scores significantly higher than DVR ones). In contrast, no differences occurred in the Body Ownership and Response subscales. 

As before, EDA scores (Figures \ref{fig:envisioning_study2_eda_embodiment}--\ref{fig:envisioning_study2_eda_end}) were computed as the sum of significant amplitudes during a 10-second window beginning 2 seconds after each event. Wilcoxon test-based analyses revealed significantly higher IVR activations at the Embodiment ($W = 137$, $p = .025$; $M_{IVR} = 0.164$, $M_{DVR} = -0.411$) and at the Intro-End events ($W = 159$, $p < .0a01$; $M_{IVR} = 0.209$, $M_{DVR} = -0.595$), while no differences occurred at the Env-Actions event ($W = 116$, $p = .24$). A Pearson's correlation between the Env-Actions EDA measurement and \emph{Response} and \emph{Body Ownership} subscales of the embodiment questionnaire showed no correlation among them (\emph{Response}: $r = -0.16$, $p = .424$; \emph{Body Ownership}: $r = -0.198$; $p = .323$).

\section{Physics Playground}
\label{sec:pp}
\subsection{Application's Design}
\label{subsec:pp_design}

\begin{figure}[t]
	\centering
	\includegraphics[width=0.8\linewidth]{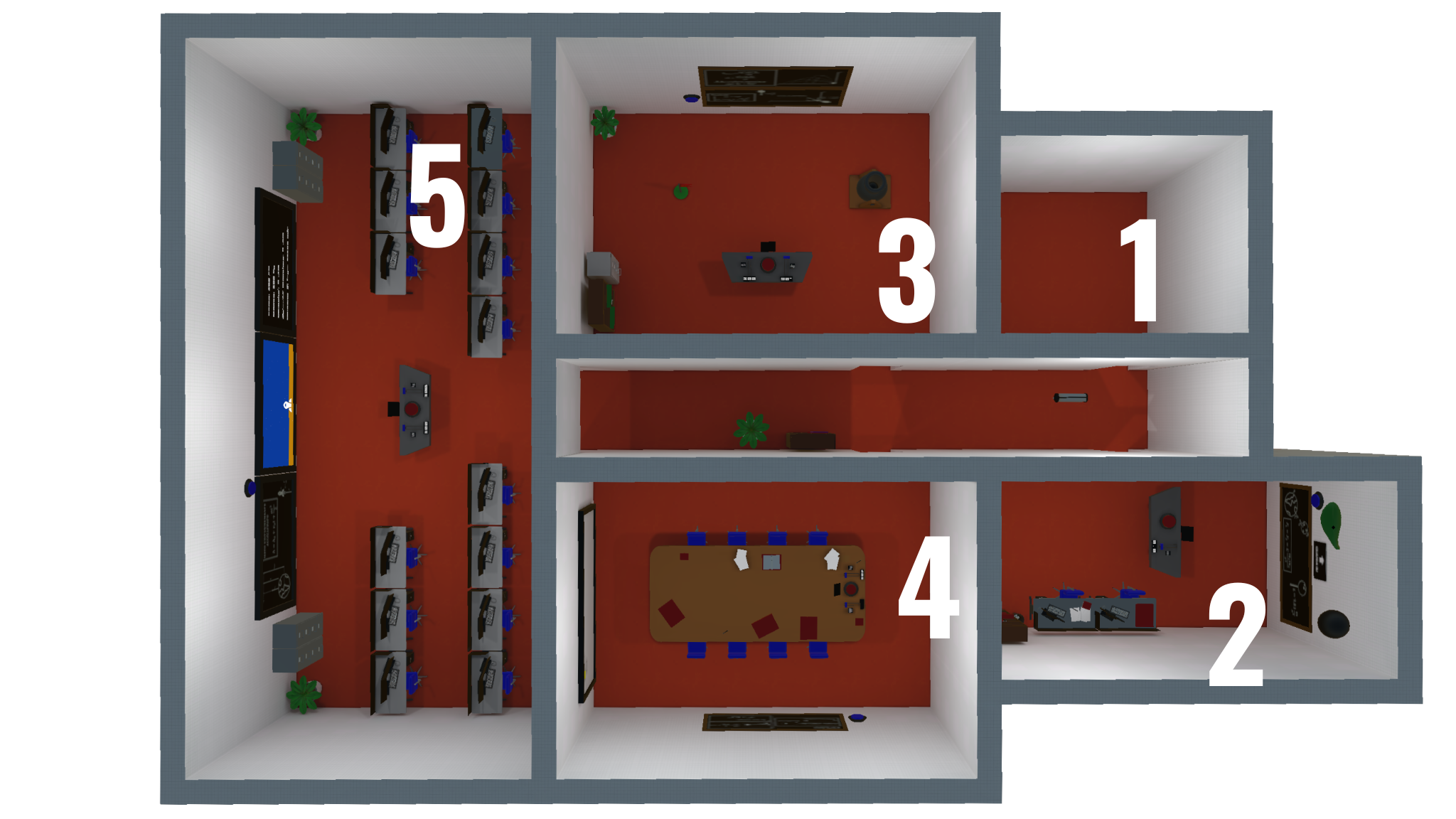}
	\caption{Top-down view of \emph{Physics Playground}'s VE, with numbers indicating the educational path to be followed.}
	\label{fig:pp-environment}
\end{figure}

\emph{Physics Playground} is designed as a virtual laboratory for exploring gravity-related phenomena through guided, hands-on experiments~\cite{battipede2025}. The user is located in an office-like VE (\autoref{fig:pp-environment}), composed of four rooms -- Gravity, Cannon, Planet, and Control -- linked by a corridor. Each room is dedicated to a core concept (free fall, projectile motion, orbital motion, escape velocity). 
Users are guided throughout the environment by a humanoid robot, Emanon, who provides theoretical explanations about the experiments and the physical variables behind them. Users can then manipulate those variables through control panels placed in each room and observe outcomes in real-time. Every room has blackboards presenting the key formulas to understand the physics laws behind the experiment better.
We included a short, dedicated tutorial area, allowing users to experiment with the interactions before entering the main environment.

\subsection{Study 1: A Mixed-Methods Evaluation}
\label{subsec:pp_study1}

The qualitative study included semi-structured interviews with professors and STEM-teaching experts and a focus group with pre-service high-school teachers, to collect insights to evaluate and optimize Physics Playground's learning experience. The discussions were guided by questions based on the Reduced Instructional Materials Motivation Survey (RIMMS) by~\citet{loorbach2015}, adapting the original material from closed to open-ended questions. 

The quantitative study was designed as a within-between subjects study, with time as the within factor (pre vs. post) and the media condition as the between factor (IVR vs. control). The DVR version was not included in this study as it aimed to preliminary evaluate the learning effectiveness of the application's design. The control condition was slideshow-based, with content mirroring the IVR and text-to-speech narration.

\subsection{Study 1: Results}
\label{subsec:pp_study1-results}

\subsubsection{Qualitative Study}

Experts and pre-service teachers generally responded positively to the experience, appreciating the application's immersive qualities and potential to engage the students, especially thanks to the possibility of interacting with simulated space environments and physics variables. Still, their opinions mainly converged on positioning \emph{Physics Playground} as a complementary, not standalone, resource, to be used after a classroom lecture (or in general after deeper theoretical introductions). Additionally, they suggested strengthening the presence of gamification elements, and personalized rather than standardized feedback. 

\subsubsection{Quantitative Study} 

Overall, $44$ people completed the study ($22$ males, $21$ females, $1$ non-binary), aged between $19$ and $34$ ($M = 23.91$, $sd = 2.74$).

\begin{figure}[t]
	\centering
	\includegraphics[width=0.7\linewidth]{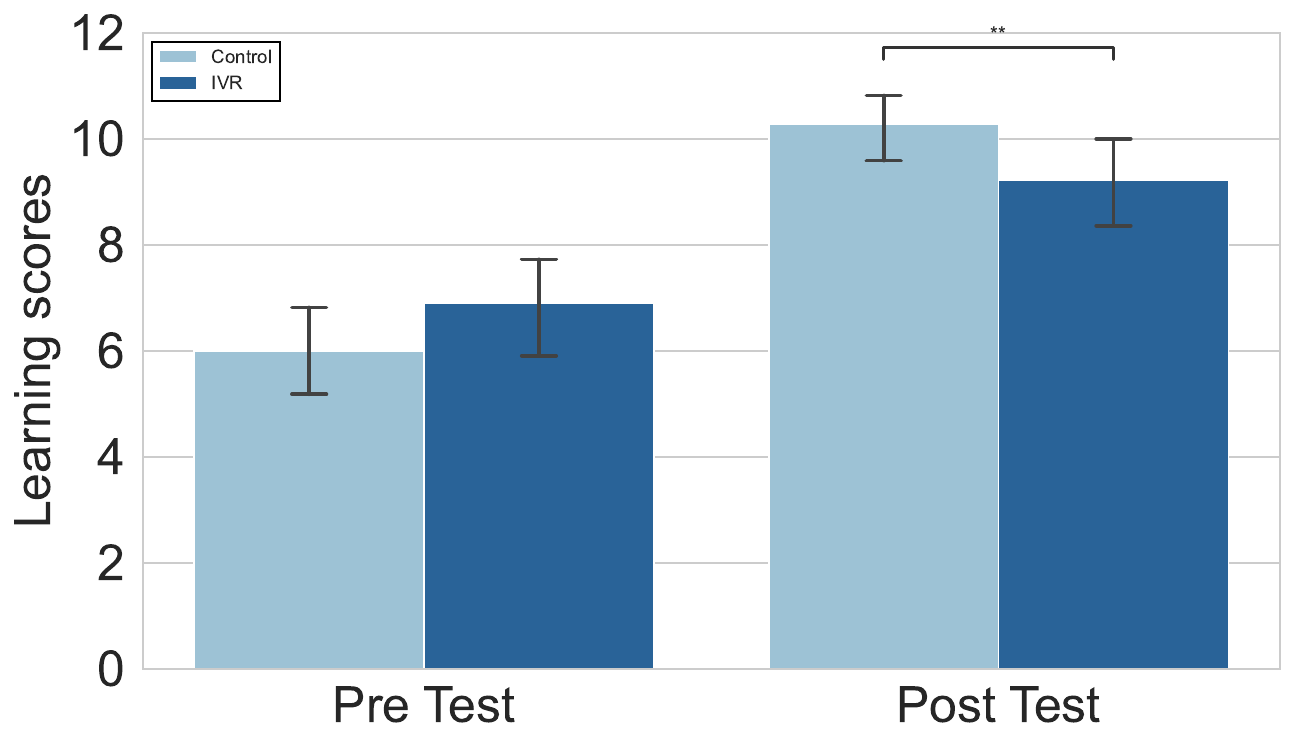}
	\caption{\emph{Physics Playground} study 1 learning results, divided by condition and time points.}
	\label{fig:pp_study1-learning}
\end{figure}

The most interesting result of this study regarded learning (displayed in \autoref{fig:pp_study1-learning}). Pre-test knowledge did not differ by condition ($t(42) = -1.38$, $p = .176$). Both IVR and control groups significantly improved their physics knowledge between pre- and post-test, as shown by a Wilcoxon test (Control: $W = 0.00$, $p < .001$, IVR: $W = 5.00$, $p < .001$). However, immediate gains favored the slideshow: a Mann-Whitney test on $\Delta_{pre{\rightarrow}post}$ showed larger gains for control ($U = 130$, $p = .008$; $M_{IVR} = 2.32$, $M_{control} = 4.27$). 

\subsection{From Study 1 to Study 2: Design and Methodological Revisions}
\label{subsec:pp_improvements}

Findings from Study 1 pointed out two complementary needs: improvements in the application's design and integration with theoretical lectures. Accordingly, \emph{Physics Playground} was revised following those indications. Some rooms were redesigned to make them more immersive and engaging, and gamification elements were included: now learners needed to complete an experiment with specific requirements to be able to proceed to the next room, with personalized feedback based on user's performance. 

Methodologically, the experimentation structure was re-framed by integrating the application with a theoretical video-lecture, introducing all the concepts presented throughout the application in detail. Fragments of the video have been embedded within the application, allowing users to review theoretical explanations before executing an experiment. 

In the second study, the slideshow control condition was replaced with an interactive web-based application, based on the PhET simulations~\cite{perkins2006}. The control condition was composed of four simulations mimicking the four rooms of the VE, which allowed users to interact with the experiments in a simple and immediate way. 

\subsection{Study 2: Ecological Evaluation}
\label{subsec:pp_study2}

To evaluate the effectiveness of the methodological and design improvements, \emph{Physics Playground} was deployed in a four-week high-school experimentation, to assess its validity and media effects in an ecological setting. The experiment was designed as a within-between subjects study, with the time (pre, immediate post, 2-week retention) as a within factor and the medium as a between factor (IVR, DVR, control). 

The experimental sequence was standardized across groups: in week 1, students completed the pre-test survey to assess their baseline physics knowledge (16 factual knowledge multiple-choice questions). In week 2, six days after the pre-test, students saw the video lecture; the day after, they participated in the assigned experimental condition and completed the post-test survey. In week 4, two weeks after the experimentation day, students answered the retention-test survey to assess knowledge retention and participated in an informal focus group to grasp qualitative insights.

In addition to learning, the collected measures included agency (with the Sense of Agency Scale questionnaire) and motivation (with the RIMMS scale).

\subsection{Study 2: Results}
\label{subsec:pp_study2-results}

Five classes participated in the study, with students randomized at the class level into IVR, DVR, and control groups. Overall, $74$ students ($43$ males, $30$ females, one non-binary) completed all the experimentation's steps and were included in the final analysis. 

\begin{figure}[t]
    \centering
\includegraphics[width=.7\linewidth]{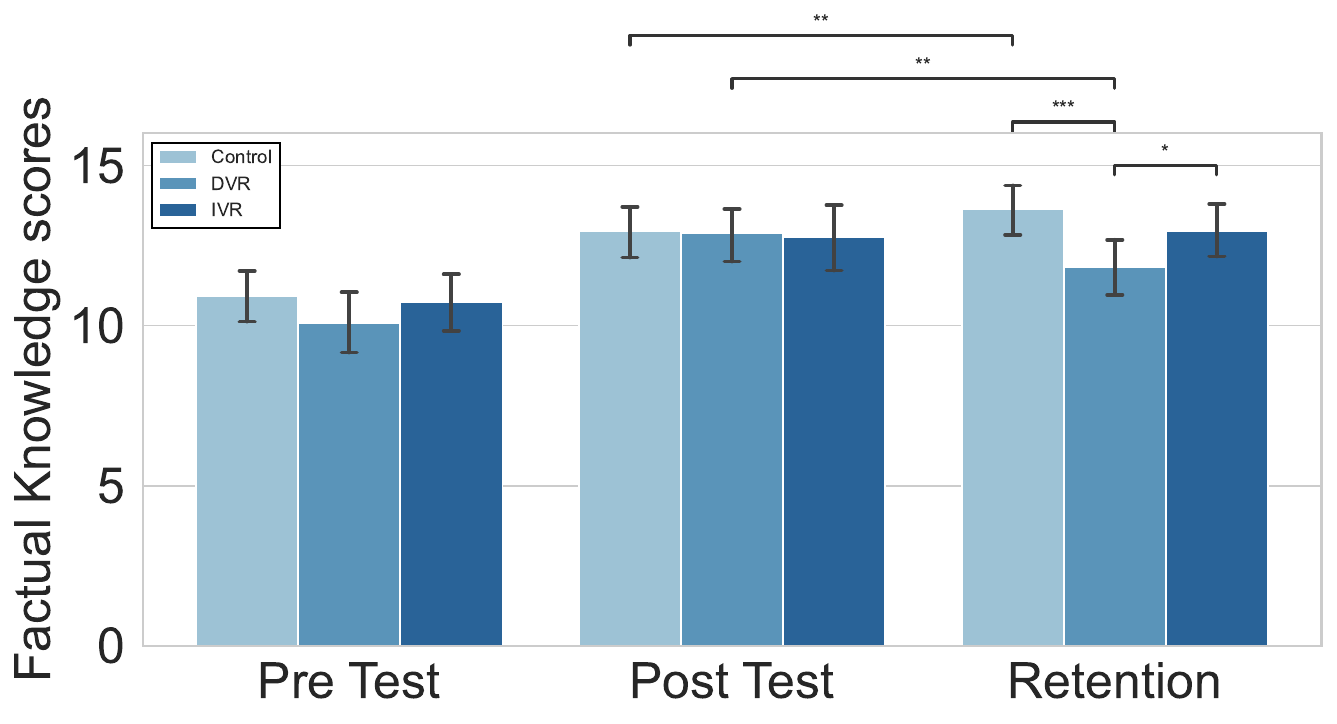}
		\caption{Factual learning results. All groups significantly improved their scores between pre- and post-test.}
		\label{subfig:pp_study2-fact-learning}
\end{figure}

Participants' scores were computed to evaluate factual knowledge acquisition by assigning one point to each correct answer and zero to the wrong ones. Pre-test scores did not differ by condition (Welch's Anova, $F = 0.951$, $p = .391$). Post-test scores were significantly higher than pre-test ones for all conditions (Student's t-test, $t = -9.6$, $p < .001$), without differences between conditions in the post-test (Welch's Anova, $F = 1.14$, $p = .328$). Analyzing retention data, a significant increase was found for the control condition ($t = -2.89$, $p = .008$) and a significant decay for DVR ($t = 2.8$, $p = .01$), with IVR scores showing no differences ($t = -0.722$, $p = .477$). Between-groups differences were found with a Welch's Anova ($F = 7.5$, $p = .002$), and post-hoc analyses revealed that IVR and control participants scored significantly higher than DVR ones (IVR vs. DVR: $p = .012$, $\Delta_{mean} = 1.24$; control vs. DVR: $p < .001$, $\Delta_{mean} = 1.71$). No differences occurred between IVR and control ($p = .521$). The results are displayed in \autoref{subfig:pp_study2-fact-learning}.

\begin{figure}[t]
    \centering
    \begin{subfigure}{0.32\linewidth}
		\includegraphics[width=\linewidth]{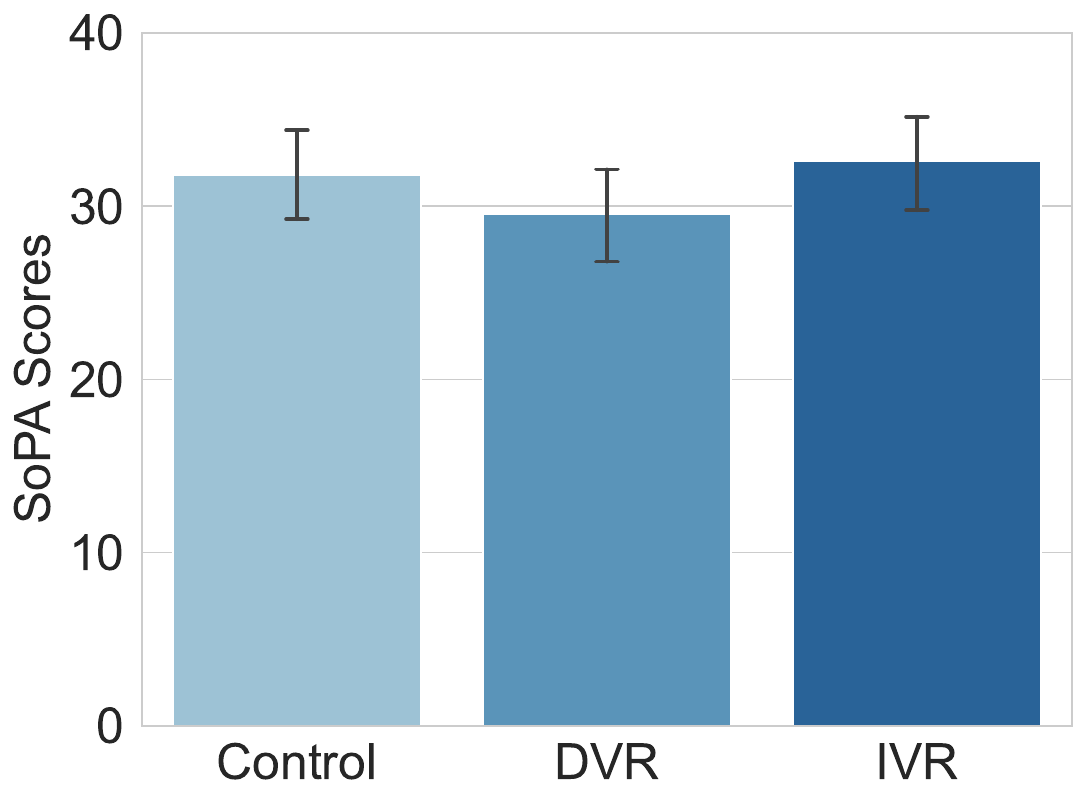}
		\caption{SoPA results.}
		\label{subfig:pp_study2-sopa}
	\end{subfigure}
\begin{subfigure}{0.32\linewidth}
		\includegraphics[width=\linewidth]{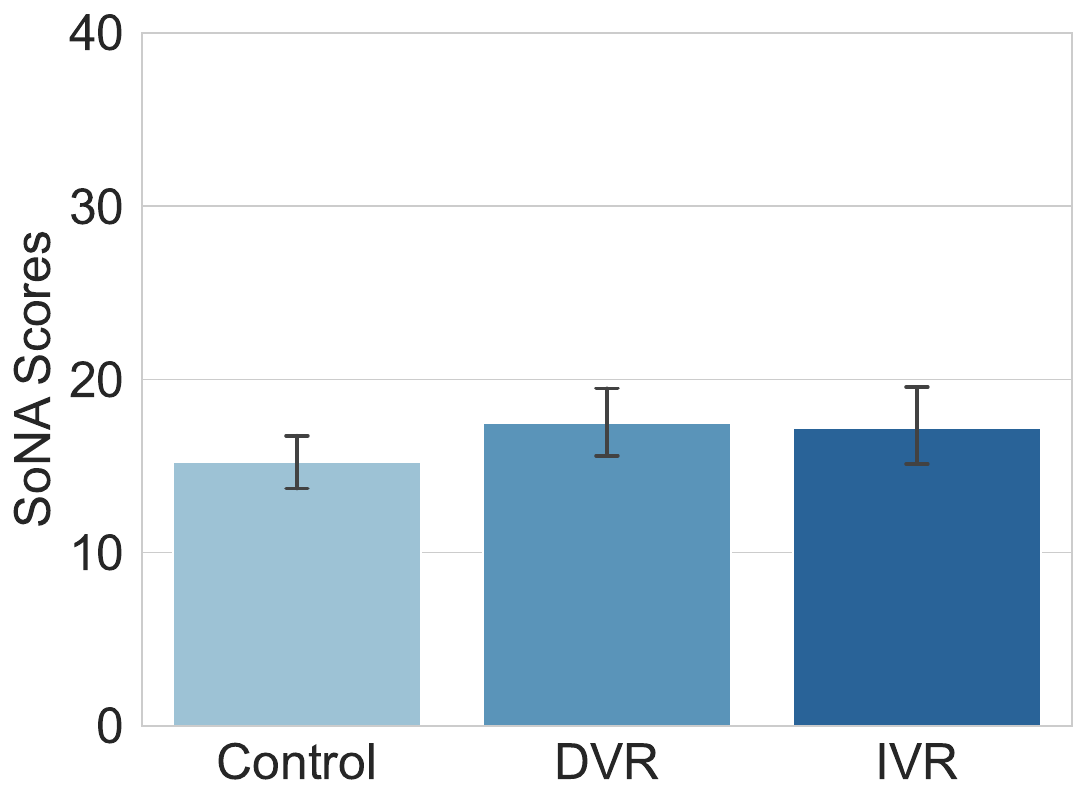}
		\caption{SoNA results.}
		\label{subfig:pp_study2-sona}
	\end{subfigure}
\begin{subfigure}{0.32\linewidth}
		\includegraphics[width=\linewidth]{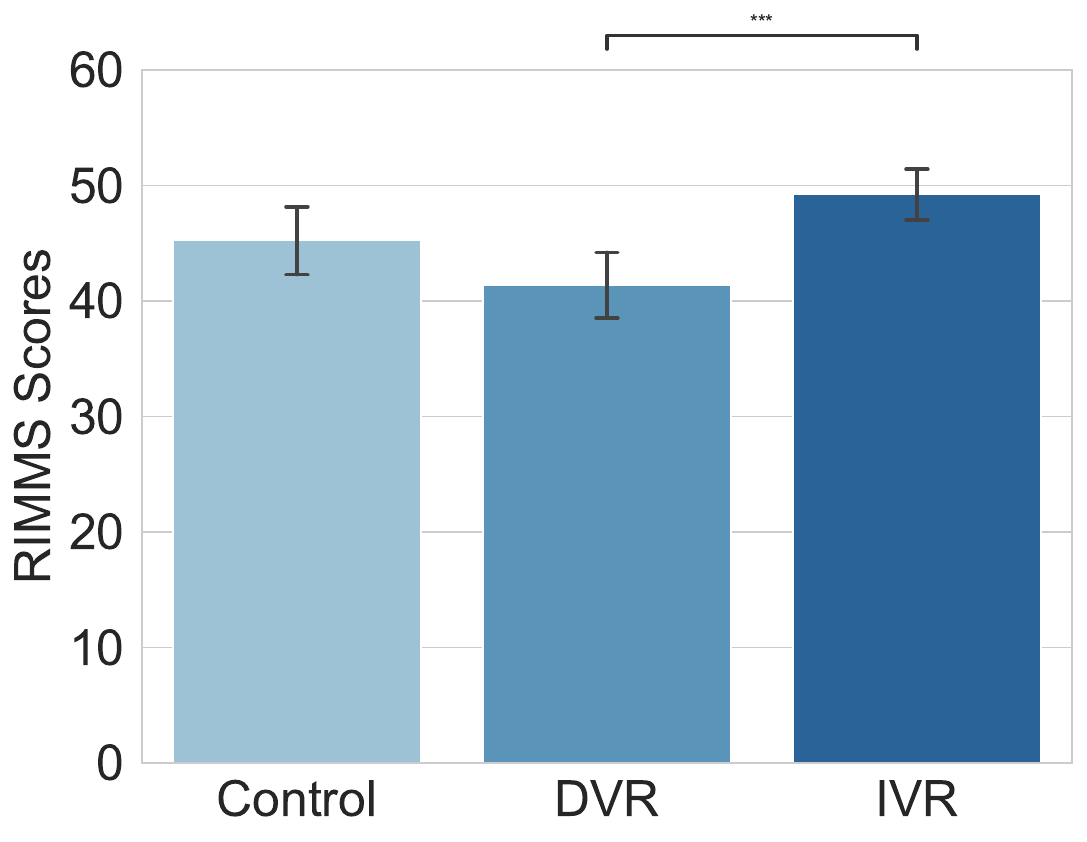}
		\caption{Motivation results}
		\label{subfig:pp_study2-motivation}
	\end{subfigure}
    \caption{\emph{Physics Playground} study 2 Agency and Motivation results, divided by condition.}
    \label{fig:pp_study2-agency-motivation}
\end{figure}

To evaluate the sense of agency, the two subscales of the Sense of Agency Scale were analyzed -- Sense of Positive Agency (SoPA; \autoref{subfig:pp_study2-sopa}) and Sense of Negative Agency (SoNA; \autoref{subfig:pp_study2-sona}). A Kruskal-Wallis test revealed no between-groups differences, neither in the SoPA ($\chi^2 = 2.11$, $p = .349$) nor in the SoNA ($\chi^2 = 2.94$, $p = .23$). 

Motivation scores were analyzed through a Kruskal-Wallis test, revealing a significant effect of condition ($\chi^2 = 14.4$, $p < .001$), with IVR significantly higher than DVR ($W = -5.65$, $p < .001$; $M_{IVR} = 49.3$, $M_{DVR} = 45.3$; \autoref{subfig:pp_study2-motivation}). No differences occurred between IVR and control ($W = 3.29$, $p = .053$) or DVR and control ($W = -2.53$, $p = .175$). 

Presence and cybersickness scores were collected on the VR groups only. On the former measure, a Mann-Whitney U test revealed higher scores for IVR than DVR ($U = 64.5$, $p < .001$; $M_{IVR} = 166$, $M_{DVR} = 134$). In addition, a Pearson's correlation between presence and agency subscales revealed a significant positive correlation with SoPA ($r = 0.468$, $p < .001$), and no correlations with SoNA ($r = -0.121$, $p = .404$). On the latter measure, a Kruskal-Wallis test found no differences, neither on the overall scale ($\chi^2 = 0.612$, $p = .434$) nor in the individual subscales (oculomotor: $\chi^2 = 0.612$, $p = .434$; disorientation: $\chi^2 = 1.354$, $p = .245$). 

Qualitative feedback were collected through an informal focus group. Students who participated consistently preferred IVR for engagement and being able to act in the VE; still, they framed it as most effective after a brief theoretical introduction, with the video lecture seen as useful preparation. DVR was viewed as a pragmatic middle ground (less immersive, easier to use), while the web-based control was appreciated for immediacy but felt less engaging. Students also reported some issues with IVR interactions, due to the novelty of the technology, highlighting how these issues will diminish with increased technology familiarity. They also strongly suggested the addition of a collaborative, multiplayer mode, to better support classroom -- and not only individual -- use.

 \section{Discussion}
\label{sec:discussion}
Triangulating across three domains, these findings show common patterns. IVR consistently improves presence and UX with respect to DVR and (when applicable) control conditions, consistent with research that links presence and sensory/interaction fidelity~\citep{witmer1998, witmer2005, slater2003}, and with reviews reporting affective and attentional gains under higher immersion~\citep{radianti2020, wu2020, hamilton2021}. At the same time, however, immediate learning gains were mixed and often media-equivalent across the studies, indicating that immersion alone does not translate directly into better near-term factual outcomes. Within CAMIL, we can argue that immersion mainly influences psychological routes, but factual knowledge acquisition does not strictly depend on it. Notably, in all the studies, cybersickness and subjective cognitive load measurements were low -- and independent of condition -- reducing confounds from external factors on learning. So, immersion per se seems not to be a strong tool to increase immediate factual knowledge gain, and effects may depend on the learning content.

Acceptance results for \emph{RHOME-VR} add details to this picture. PEU did not differ by medium, while PU was higher in IVR. In practice, participants experienced the immersive tour as at least as easy to use and more useful than traditional instructional tools. Within CAMIL, this pattern is coherent with psychological routes: stronger presence and affect (including a plausible novelty boost) can increase usefulness perception, even if the learning outcomes did not differ. This ``innovative but not harder'' profile is encouraging for IVR adoption. 

Immersion also seems to give learning advantages when dealing with longer-term retention. In \emph{Physics Playground}, IVR participants retained more than DVR ones, showing how immersion provides additional value in retaining knowledge when dealing with complex, 3D interactive applications. Analyzing the literature, we noticed mixed results on this topic, with some papers showing an advantage of immersion in knowledge retention~\citep{raya2018, lai2022} and others with either no differences~\citep{buttussi2018, makransky2019a, klingenberg2020, madden2020} or even with DVR superior to IVR~\cite{smith2018}. Moreover, a simple interactive 2D control condition in \emph{Physics Playground}'s study 2 had comparable results with IVR. 

These findings raise three points. First, the data suggest that engagement and interactivity, rather than technological complexity per se, are central to knowledge retention: when paired with active manipulation of dynamic phenomena, IVR supported better two-week recall than DVR; yet a simple, interactive 2D simulation performed similarly -- pointing to the importance of an active interaction loop aligned to the target concept. Second, in line with CAMIL, these effects appear when task-media fit is high -- so when the medium best fits the task for the user -- and extraneous load is limited; this helps explain why representation-dense, labeling tasks in \emph{RHOME-VR} did not benefit at post-test from higher immersion, whereas hands-on exploration in \emph{Physics Playground} did at follow-up. Third, instructional integration matters: adding a theoretical lecture before the immersive experience improved immediate results over prior studies and supported retention, suggesting blended learning -- brief pre-training or theoretical lectures plus digital, interactive hands-on sessions -- as a promising strategy. Still, to the best of our knowledge, few works adopt this approach in educational VR research, and findings are mixed (IVR $>$ DVR~\cite{ding2024}; no differences~\citep{lai2022, checa2023}), revealing that further research is needed to understand the actual contribution of this approach better.

Other observations emerged around agency. In \emph{Physics Playground}, Sense of Agency scores did not differ by medium in either SoPA or SoNA subscales. Significantly, presence correlated positively with SoPA -- that is, the more participants felt ``being there'', the more they actually felt in control within the simulation -- while SoNA showed no relationship, suggesting that higher presence does not automatically reduce feelings of diminished control~\cite{ortiz2023}. From these results, we can provide possible interpretations. First, agency is not a single thing: classroom studies distinguish agency over actions (motor control), over attention (where to look), and over learning (what to pursue), and show that making an experience more interactive mainly boosts action-level agency, not necessarily agency over learning or focus~\cite{mcgivney2025}. Moreover, even low-interactivity media (such as immersive video) can support agency depending on the design goal and learner identity~\cite{mcgivney2025}. So, the device alone is not determinative: media design and learner characteristics can moderate agency. This helps explain why SoPA did not differ by medium even as presence rose: clear, responsive controls and autonomy can make 2D and DVR participants feel agentic. At the same time, complex IVR may reduce this feeling if interaction complexity is high. These observations are also reflected by qualitative feedback from the focus group: students highlighted how the novelty of the VR -- and mainly of the IVR -- approach overwhelmed them sometimes, given their unfamiliarity with the technology and the controls. Pre-training and longer media exposure may reduce this issue, leading to higher perceived agency in IVR and, ultimately, reduced confusion and higher learning outcomes.  

The \emph{Envisioning Corals} results complement this picture. Here, embodiment and agency results were higher in IVR than in DVR, consistent with a design that proposed natural interactions rather than mouse and keyboard mediation. Behavioral logs add an important nuance: more locomotion correlated with lower agency. A likely driver in this result is the difference in interaction and movement modalities across media. In fact, IVR participants could visually inspect the scene by simply turning their heads and moving via teleportation, with potentially lower effort required and more efficient movement strategies applicable. On the other hand, DVR participants had to rely on mouse-based observations and continuous movement via WASD pressure. In CAMIL terms, the IVR setup reduces control overhead and preserves agency for the task (collecting food, avoiding debris), whereas DVR participants may have been forced to dedicate attentional resources to wayfinding. 

Beyond media differences, avatar choice mattered in \emph{Envisioning Corals}. In the embodiment study, overall embodiment differed by avatar, and physiological activations tracked those differences. Read through CAMIL, this points to both the identity route (how much the user can \emph{be} the agent) and the control route (how well the body schema and affordances align with the required actions). Avatars whose morphology maps more naturally onto the user's sensorimotor characteristics can strengthen embodiment and arousal without adding load. 

Other interesting results were presented in \emph{Envisioning Corals} when dealing with EDA. Activation was higher in IVR at the beginning of the session, consistent with prior research showing that immersion amplifies emotional engagement~\citep{kuhnel2023, niu2019, beck2018}. However, this difference disappeared at the end (just before the pro-environmental guidance audio). Read through CAMIL, immersion appears to boost the initial affective route, while the narrative and environmental design equalize engagement across media. In practice, the early IVR arousal window can generate attention and curiosity; after a longer exposure, though, content takes precedence over technology, leading to comparable emotional arousal. This helps us explain why learning outcomes in \emph{Envisioning Corals} were media-equivalent despite different early arousal profiles. This point is mirrored in the embodiment results: the \emph{Response} and \emph{Body Ownership} subscales did not differ between IVR and DVR, possibly due to VE and content design. Neither subscale correlated with EDA at the final event, suggesting a decoupling between subjective embodiment and emotional arousal. What caused these results remains an open question for future investigation.
 
\section{Conclusions and Design Implications}
\label{sec:conclusions}
This work examined when and how immersion helps in education across three domains -- cultural heritage (\emph{RHOME-VR}), environmental awareness (\emph{Envisioning Corals}), and physics (\emph{Physics Playground}) -- under a shared methodological backbone. Consistent with prior research~\cite{wu2020, radianti2020, hamilton2021}, IVR consistently improved presence and UX, while immediate factual learning remained media-equivalent. Where IVR did help learning was in longer-term retention: in \emph{Physics Playground}, IVR outperformed DVR at the two-week test, even if a simple interactive 2D control had comparable retention results, suggesting that active manipulation and task-media fit -- not technological complexity alone -- drive long-term retention. Agency findings further underscored the role of control factors: clear, responsive interactions supported a sense of positive agency across media; in \emph{Envisioning Corals}, natural interaction and low-overhead locomotion boosted embodiment/agency, whereas greater locomotion demands correlated with lower agency. Physiological data showed an early arousal advantage for IVR, lost later as narrative/content took precedence over technology. 

These results suggest some design indications for instructional content. First, instructors should prioritize IVR for retention-oriented goals when dealing with 3D manipulations. In fact, using IVR when learners must manipulate 3D objects and observe immediate consequences can strengthen consolidation over time. 

Second, our results suggest that IVR -- as noted also by previous literature~\cite{wu2020} -- should be blended with theoretical introductions. The classroom deployment indicates that a short video lecture improves near-term learning performance, thereby increasing the learning effectiveness of the technology.

Third, avatar choice and interaction mappings should feel natural. In line with the embodiment results, morphologies and input strategies that align with learners' sensorimotor expectations elevate agency and embodiment without adding load. Where appropriate, hand tracking can further increase naturalness; however, it removes controller haptics and may reduce grasp confirmations. Designers should weigh this trade-off against task needs (e.g., whether tactile feedback is pedagogically helpful or visual/auditory confirmations suffice). Locomotion should likewise minimize control overhead: head-gaze and teleportation supported efficient scene exploration in IVR, whereas continuous keyboard locomotion pushed attention toward wayfinding rather than task goals.

Fourth, IVR's early arousal window should be leveraged to focus attention. Physiological data showed stronger initial engagement under IVR that later converged across media as narrative and task content took precedence. Structure sessions so the opening minutes capture attention (clear goals, brief challenge, immediate cause-and-effect), then let content and hands-on activity carry learning once arousal levels even out. Keeping locomotion light and interactions direct in this phase preserves attentional resources for the concept rather than the interface.

Finally, multiple IVR exposures should be planned to temper novelty and build control fluency. Sense of agency findings and student feedback suggest unfamiliarity with devices and mappings can suppress perceived control, particularly on first exposure. Short, repeated sessions --paired with pre-use tutorials and consistent mappings across modules -- raise positive agency, reduce confusion, and help translate motivational advantages into stable outcomes. These recommendations extend our conclusion: IVR is most effective when its immersive affordances are matched to 3D, action-centric objectives and delivered through natural, low-overhead control designs that capitalize on early engagement while building durable skill over repeated use.

\subsection{Limitations \& Future Works}

These findings should be interpreted in light of some limitations that may have influenced them.

Across applications, factual knowledge was primarily assessed as a learning outcome. There was no systematic evaluation of conceptual or procedural knowledge, which may be differentially influenced by immersion, presence, and control factors. 
Future work should evaluate different kinds of knowledge, including a deeper and more robust evaluation of transfer of learning.

Most analyses targeted immediate learning, and where retention was included, the interval was two weeks. While this afforded an ecological check in high schools, it limits claims about the longer-term durability of media effects. Longer (e.g., 1--3 months) or multi-step follow-ups (e.g., after one week, one month, and six months) are needed to characterize consolidation better.

Many constructs -- presence, motivation, agency, acceptance -- were measured through self-reported measures. Although standard in the literature, self-reports are sensitive to memory, demand, and expectancy effects and do not fully capture cognitive load or affective dynamics. Only \emph{Envisioning Corals} incorporated physiology (EDA), but the analysis focused on short windows around predefined events. A richer, multi-method battery would improve construct validity. 

In \emph{RHOME-VR}, IVR was run in-lab while DVR/control were remote, introducing a context confound for UX/TAM/presence despite content parity. In \emph{Physics Playground}'s study 2, classes were randomized at the class level; our primary analyses did not model clustering, hence intra-class correlation may have biased standard errors. Future replications should use mixed-effects models (random intercepts for class/teacher) or cluster-robust procedures. Additionally, cross-media interaction/locomotion could not be fully harmonized (\emph{Envisioning Corals}: teleport + head-gaze in IVR vs. continuous WASD + mouse-look in DVR). While this reflects realistic affordances, it also makes it harder to attribute differences purely to `immersion' rather than control overhead. Matched locomotion and feedback should be used to isolate media effects properly. 

Many participants were first-time IVR users. Novelty can both elevate motivation/PU and temporarily depress perceived agency until controls are mastered, potentially increasing acceptance yet masking learning/control benefits~\cite{huang2003}. Longitudinal designs with repeated IVR exposures are needed to limit novelty's impact and properly evaluate the technological effects.

These limitations motivate several research directions. First, exploring learning outcomes beyond factual recall to include conceptual and procedural, with transfer tasks redesigned to minimize practice and effort confounds. Second, pair self-reports with multimodal signals and behavioral traces (e.g., HRV/EDA/eye movement plus in-app actions) and use them to model CAMIL routes more directly.

On the instructional-design front, test generative learning strategies (GLS)~\cite{GLS_FiorellaMayer2015} inside IVR (e.g., in-situ self-explanations, prompted comparisons) to examine whether immersion that elicits generation -- as opposed to exposure -- improves near- and long-term outcomes. Personalization is another promising lever: learner personas~\cite{tauro2022personas} or lightweight AI profiling~\cite{souza2024} could tune avatar morphology, control mappings, and pacing to the user's goals and abilities, potentially strengthening CAMIL's identity and control routes. Social dimensions also merit study: collaborative IVR with shared tasks and communication tools may alter presence/agency dynamics and outcomes relative to solo use.

Finally, we encourage ecological evaluations where IVR is integrated into a broader didactic path (pre-training, IVR lab, post-activities), with repeated exposures to mitigate novelty and build control fluency. Adopting such an experimental setup would allow us to better understand the impact of this technology in the actual environment where it (potentially) could be used in the future.



\begin{thebibliography}{97}
	\providecommand{\natexlab}[1]{#1}
	\providecommand{\url}[1]{#1}
	\csname url@samestyle\endcsname
	\providecommand{\newblock}{\relax}
	\providecommand{\bibinfo}[2]{#2}
	\providecommand{\BIBentrySTDinterwordspacing}{\spaceskip=0pt\relax}
	\providecommand{\BIBentryALTinterwordstretchfactor}{4}
	\providecommand{\BIBentryALTinterwordspacing}{\spaceskip=\fontdimen2\font plus
		\BIBentryALTinterwordstretchfactor\fontdimen3\font minus
		\fontdimen4\font\relax}
	\providecommand{\BIBforeignlanguage}[2]{{%
			\expandafter\ifx\csname l@#1\endcsname\relax
			\typeout{** WARNING: IEEEtranN.bst: No hyphenation pattern has been}%
			\typeout{** loaded for the language `#1'. Using the pattern for}%
			\typeout{** the default language instead.}%
			\else
			\language=\csname l@#1\endcsname
			\fi
			#2}}
	\providecommand{\BIBdecl}{\relax}
	\BIBdecl
	
	\bibitem[Jensen and Konradsen(2018)]{jensen-konradsen2018}
	L.~Jensen and F.~Konradsen, ``A review of the use of virtual reality
	head-mounted displays in education and training,'' \emph{Educ Inf Technol},
	vol.~23, pp. 1515--1529, 2018.
	
	\bibitem[Radianti et~al.(2020)Radianti, Majchrzak, Fromm, and
	Wohlgenannt]{radianti2020}
	J.~Radianti, T.~A. Majchrzak, J.~Fromm, and I.~Wohlgenannt, ``A systematic
	review of immersive virtual reality applications for higher education: Design
	elements, lessons learned, and research agenda,'' \emph{Computers \&
		Education}, vol. 147, p. 103778, 2020.
	
	\bibitem[Di~Natale et~al.(2020)Di~Natale, Repetto, Riva, and
	Villani]{dinatale2020}
	A.~F. Di~Natale, C.~Repetto, G.~Riva, and D.~Villani, ``Immersive virtual
	reality in k-12 and higher education: A 10-year systematic review of
	empirical research,'' \emph{British Journal of Educational Technology},
	vol.~51, no.~6, pp. 2006--2033, 2020.
	
	\bibitem[Buttussi and Chittaro(2018)]{buttussi2018}
	F.~Buttussi and L.~Chittaro, ``Effects of different types of virtual reality
	display on presence and learning in a safety training scenario,'' \emph{IEEE
		Transactions on Visualization and Computer Graphics}, vol.~24, no.~2, pp.
	1063--1076, 2018.
	
	\bibitem[Merchant et~al.(2014)Merchant, Goetz, Cifuentes, Keeney-Kennicutt, and
	Davis]{merchant2014}
	Z.~Merchant, E.~T. Goetz, L.~Cifuentes, W.~Keeney-Kennicutt, and T.~J. Davis,
	``Effectiveness of virtual reality-based instruction on students' learning
	outcomes in k-12 and higher education: A meta-analysis,'' \emph{Computers \&
		Education}, vol.~70, pp. 29--40, 2014.
	
	\bibitem[Pellas et~al.(2021)Pellas, Mystakidis, and Kazanidis]{pellas2021}
	N.~Pellas, S.~Mystakidis, and I.~Kazanidis, ``{Immersive Virtual Reality in
		K-12 and Higher Education: A systematic review of the last decade scientific
		literature},'' \emph{Virtual Reality}, vol.~25, pp. 835--861, 2021.
	
	\bibitem[Wu et~al.(2020)Wu, Yu, and Gu]{wu2020}
	B.~Wu, X.~Yu, and X.~Gu, ``Effectiveness of immersive virtual reality using
	head-mounted displays on learning performance: A meta-analysis,''
	\emph{British Journal of Educational Technology}, vol.~51, no.~6, pp.
	1991--2005, 2020.
	
	\bibitem[Makransky et~al.(2019{\natexlab{a}})Makransky, Terkildsen, and
	Mayer]{makransky2019b}
	G.~Makransky, T.~S. Terkildsen, and R.~E. Mayer, ``Adding immersive virtual
	reality to a science lab simulation causes more presence but less learning,''
	\emph{Learning and Instruction}, vol.~60, pp. 225--236, 2019.
	
	\bibitem[Makransky and Petersen(2021)]{makransky2021camil}
	G.~Makransky and G.~B. Petersen, ``The cognitive affective model of immersive
	learning (camil): a theoretical research-based model of learning in immersive
	virtual reality,'' \emph{Educational Psychology Review}, vol.~33, no.~3, pp.
	937--958, 2021.
	
	\bibitem[Paino~Ambrosio and Rodriguez~Fidalgo(2020)]{ambrosiofidalgo2020}
	A.~Paino~Ambrosio and M.~I. Rodriguez~Fidalgo, ``Past, present and future of
	virtual reality: Analysis of its technological variables and definitions,''
	\emph{Culture \& History Digital Journal}, pp. 10--2253, 06 2020.
	
	\bibitem[Biocca and Delaney(1995)]{biocca1995}
	F.~Biocca and B.~Delaney, \emph{Immersive virtual reality technology}.\hskip
	1em plus 0.5em minus 0.4em\relax USA: L. Erlbaum Associates Inc., 1995, pp.
	57--124.
	
	\bibitem[Chavez and Bayona(2018)]{chavez2018}
	B.~Chavez and S.~Bayona, ``Virtual reality in the learning process,'' in
	\emph{Trends and Advances in Information Systems and Technologies}, 2018.
	
	\bibitem[Lee and Wong(2014)]{lee2014}
	E.~A.-L. Lee and K.~W. Wong, ``Learning with desktop virtual reality: low
	spatial ability learners are more positively affected,'' \emph{Computers \&
		Education}, vol.~79, pp. 49--58, 2014.
	
	\bibitem[Walsh and Pawlowski(2002)]{walsh2002}
	K.~Walsh and S.~Pawlowski, ``Virtual reality: A technology in need of is
	research,'' \emph{Communications of the Association for Information Systems},
	vol.~8, 2002.
	
	\bibitem[Ryan(2015)]{ryan2015}
	M.-L. Ryan, \emph{Narrative as a virtual reality 2: Revisiting immersion and
		interactivity in literature and electronic media}.\hskip 1em plus 0.5em minus
	0.4em\relax JHU press, 2015.
	
	\bibitem[Steuer(1995)]{steuer1995}
	J.~Steuer, ``Defining virtual reality: Dimensions determining telepresence,''
	\emph{Communication in the age of virtual reality}, vol.~42, pp. 73--93,
	1995.
	
	\bibitem[Johnson et~al.(2022)Johnson, Bailey, Schroeder, and
	Marraffino]{johnson2022}
	C.~Johnson, S.~Bailey, B.~Schroeder, and M.~Marraffino, ``Procedural learning
	in virtual reality: The role of immersion, interactivity, and spatial
	ability.'' \emph{Technology, Mind, and Behavior}, vol.~3, 11 2022.
	
	\bibitem[Velasco and Obrist(2021)]{velasco2021}
	C.~Velasco and M.~Obrist, ``Multisensory experiences: A primer,'' \emph{Front.
		Comput. Sci.}, vol.~3, p. 614524, 2021.
	
	\bibitem[Skarbez et~al.(2017)Skarbez, Brooks, and Whitton]{skarbez2017}
	R.~Skarbez, F.~P. Brooks, Jr., and M.~C. Whitton, ``A survey of presence and
	related concepts,'' \emph{ACM Comput. Surv.}, vol.~50, no.~6, 2017.
	
	\bibitem[Witmer and Singer(1998)]{witmer1998}
	B.~G. Witmer and M.~J. Singer, ``Measuring presence in virtual environments: A
	presence questionnaire,'' \emph{Presence: Teleoperators and Virtual
		Environments}, vol.~7, no.~3, pp. 225--240, 1998.
	
	\bibitem[Nilsson and Serafin(2016)]{nilsson2016}
	N.~Nilsson and S.~Serafin, ``Immersion revisited: A review of existing
	definitions of immersion and their relation to different theories of
	presence,'' \emph{Human Technology}, vol.~12, pp. 108--134, 11 2016.
	
	\bibitem[Witmer et~al.(2005)Witmer, Jerome, and Singer]{witmer2005}
	B.~G. Witmer, C.~J. Jerome, and M.~J. Singer, ``The factor structure of the
	presence questionnaire,'' \emph{Presence}, vol.~14, no.~3, pp. 298--312,
	2005.
	
	\bibitem[Slater et~al.(1994)Slater, Usoh, and Steed]{slater1994}
	M.~Slater, M.~Usoh, and A.~Steed, ``Depth of presence in virtual
	environments,'' \emph{Presence: Teleoper. Virtual Environ.}, vol.~3, no.~2,
	pp. 130--144, 1994.
	
	\bibitem[Slater(2003)]{slater2003}
	M.~Slater, ``A note on presence terminology,'' Online, 2003, [Online].
	Available at \url{https://tinyurl.com/slaterpresenceterminology2003}.
	
	\bibitem[Slater(2018)]{slater2018}
	------, ``Immersion and the illusion of presence in virtual reality,'' \emph{Br
		J Psychol.}, vol.~3, no. 109, pp. 431--433, 2018.
	
	\bibitem[Cipresso et~al.(2018)Cipresso, Giglioli, Raya, and Riva]{cipresso2018}
	P.~Cipresso, I.~Giglioli, M.~Raya, and G.~Riva, ``The past, present, and future
	of virtual and augmented reality research: A network and content analysis of
	articles published in frontiers in psychology,'' \emph{Frontiers in
		Psychology}, vol.~9, p. 2086, 2018.
	
	\bibitem[Kami\'{n}ska et~al.(2019)Kami\'{n}ska, Sapiński, Wiak, Tikk, Haamer,
	Avots, Helmi, Ozcinar, and Anbarjafari]{kaminska2019}
	D.~Kami\'{n}ska, T.~Sapiński, S.~Wiak, T.~Tikk, R.~E. Haamer, E.~Avots,
	A.~Helmi, C.~Ozcinar, and G.~Anbarjafari, ``Virtual reality and its
	applications in education: Survey,'' \emph{Information}, vol.~10, no.~10,
	2019.
	
	\bibitem[Christou(2010)]{christou2010}
	C.~Christou, \emph{Virtual Reality in Education}, 2010, pp. 228--243.
	
	\bibitem[Wilson(2002)]{wilson2002}
	M.~Wilson, ``Six views of embodied cognition,'' \emph{Psychonomic Bulletin \&
		Review}, vol.~9, pp. 625--636, 2002.
	
	\bibitem[Ionescu and Vasc(2014)]{ionescu2014}
	T.~Ionescu and D.~Vasc, ``Embodied cognition: Challenges for psychology and
	education,'' \emph{Procedia - Social and Behavioral Sciences}, vol. 128, pp.
	275--280, 2014, international Conference: EDUCATION AND PSYCHOLOGY CHALLENGES
	- TEACHERS FOR THE KNOWLEDGE SOCIETY -- 2nd EDITION EPC -- TKS 2013.
	
	\bibitem[Lindgren and Johnson-Glenberg(2013)]{lindgren2013}
	R.~Lindgren and M.~Johnson-Glenberg, ``Emboldened by embodiment: Six precepts
	for research on embodied learning and mixed reality,'' \emph{Educational
		Researcher}, vol.~42, no.~8, pp. 445--452, 2013.
	
	\bibitem[Johnson-Glenberg et~al.(2021)Johnson-Glenberg, Bartolomea, and
	Kalina]{johnson-glenberg2021}
	M.~C. Johnson-Glenberg, H.~Bartolomea, and E.~Kalina, ``Platform is not
	destiny: Embodied learning effects comparing 2d desktop to 3d virtual reality
	stem experiences,'' \emph{Journal of Computer Assisted Learning}, vol.~37,
	no.~5, pp. 1263--1284, 2021.
	
	\bibitem[Mayer(2014)]{mayer2014}
	R.~E. Mayer, \emph{Cognitive Theory of Multimedia Learning}, ser. Cambridge
	Handbooks in Psychology.\hskip 1em plus 0.5em minus 0.4em\relax Cambridge
	University Press, 2014, pp. 43--71.
	
	\bibitem[Anderson and Krathwohl(2001)]{anderson2001}
	L.~W. Anderson and D.~R. Krathwohl, \emph{A taxonomy for learning, teaching,
		and assessing: A revision of Bloom's taxonomy of educational objectives:
		complete edition}.\hskip 1em plus 0.5em minus 0.4em\relax Addison Wesley
	Longman, Inc., 2001.
	
	\bibitem[Parong et~al.(2020)Parong, Pollard, Files, Oiknine, Sinatra, Moss,
	Passaro, and Khooshabeh]{parong2020}
	J.~Parong, K.~A. Pollard, B.~T. Files, A.~H. Oiknine, A.~M. Sinatra, J.~D.
	Moss, A.~Passaro, and P.~Khooshabeh, ``The mediating role of presence differs
	across types of spatial learning in immersive technologies,'' \emph{Computers
		in Human Behavior}, vol. 107, p. 106290, 2020.
	
	\bibitem[Pollard et~al.(2020)Pollard, Oiknine, Files, Sinatra, Patton, Ericson,
	Thomas, and Khooshabeh]{pollard2020}
	K.~A. Pollard, A.~H. Oiknine, B.~T. Files, A.~M. Sinatra, D.~Patton,
	M.~Ericson, J.~Thomas, and P.~Khooshabeh, ``Level of immersion affects
	spatial learning in virtual environments: results of a three-condition
	within-subjects study with long intersession intervals,'' \emph{Virtual
		Reality}, vol.~24, pp. 783--796, 2020.
	
	\bibitem[{De Witte} et~al.(2024){De Witte}, Reynaert, Hutain, Kieken, Jabbour,
	and Possik]{dewitte2024}
	B.~{De Witte}, V.~Reynaert, J.~Hutain, D.~Kieken, J.~Jabbour, and J.~Possik,
	``Immersive learning of factual knowledge while assessing the influence of
	cognitive load and spatial abilities,'' \emph{Computers \& Education: X
		Reality}, vol.~5, p. 100085, 2024.
	
	\bibitem[Thompson et~al.(2021)Thompson, Uz-Bilgin, Tutwiler, Anteneh, Meija,
	Wang, Tan, Eberhardt, Roy, and Klopfer]{thompson2021}
	M.~Thompson, C.~Uz-Bilgin, M.~Tutwiler, M.~Anteneh, J.~Meija, A.~Wang, P.~Tan,
	R.~Eberhardt, D.~Roy, and E.~Klopfer, ``Immersion positively affects learning
	in virtual reality games compared to equally interactive 2d games,''
	\emph{Information and Learning Sciences}, vol. ahead-of-print, 2021.
	
	\bibitem[Araiza-Alba et~al.(2021)Araiza-Alba, Keane, Chen, and
	Kaufman]{araiza-alba2021}
	P.~Araiza-Alba, T.~Keane, W.~S. Chen, and J.~Kaufman, ``Immersive virtual
	reality as a tool to learn problem-solving skills,'' \emph{Computers \&
		Education}, vol. 164, p. 104121, 2021.
	
	\bibitem[Tang et~al.(2022)Tang, Wang, Liu, Liu, and Jiang]{tang2022}
	Q.~Tang, Y.~Wang, H.~Liu, Q.~Liu, and S.~Jiang, ``Experiencing an art education
	program through immersive virtual reality or ipad: Examining the mediating
	effects of sense of presence and extraneous cognitive load on enjoyment,
	attention, and retention,'' \emph{Frontiers in Psychology}, vol.~13, 2022.
	
	\bibitem[Checa et~al.(2023)Checa, Miguel-Alonso, and Bustillo]{checa2023}
	D.~Checa, I.~Miguel-Alonso, and A.~Bustillo, ``Immersive virtual-reality
	computer-assembly serious game to enhance autonomous learning,''
	\emph{Virtual Reality}, vol.~27, pp. 3301--3318, 2023.
	
	\bibitem[Madden et~al.(2020)Madden, Pandita, Schuldt, Kim, S.~Won, and
	Holmes]{madden2020}
	J.~Madden, S.~Pandita, J.~P. Schuldt, B.~Kim, A.~S.~Won, and N.~G. Holmes,
	``Ready student one: Exploring the predictors of student learning in virtual
	reality,'' \emph{PLOS ONE}, vol.~15, no.~3, pp. 1--26, 2020.
	
	\bibitem[Petersen et~al.(2022)Petersen, Petkakis, and Makransky]{petersen2022}
	G.~B. Petersen, G.~Petkakis, and G.~Makransky, ``A study of how immersion and
	interactivity drive vr learning,'' \emph{Computers \& Education}, vol. 179,
	p. 104429, 2022.
	
	\bibitem[Kaplan-Rakowski et~al.(2024)Kaplan-Rakowski, Cockerham, and
	Ferdig]{kaplan-rakowski2024}
	R.~Kaplan-Rakowski, D.~Cockerham, and R.~E. Ferdig, ``The impact of sound and
	immersive experience on learners when using virtual reality and tablet: A
	mixed-method study,'' \emph{British Journal of Educational Technology},
	vol.~55, no.~4, pp. 1560--1582, 2024.
	
	\bibitem[Mulders(2023)]{mulders2023}
	M.~Mulders, ``Learning about victims of holocaust in virtual reality: The main,
	mediating and moderating effects of technology, instructional method, flow,
	presence, and prior knowledge,'' \emph{Multimodal Technologies and
		Interaction}, vol.~7, no.~3, 2023.
	
	\bibitem[Cecil et~al.(2018)Cecil, Gupta, Pirela-Cruz, and
	Ramanathan]{cecil2018}
	J.~Cecil, A.~Gupta, M.~Pirela-Cruz, and P.~Ramanathan, ``A network-based
	virtual reality simulation training approach for orthopedic surgery,''
	\emph{ACM Trans. Multimedia Comput. Commun. Appl.}, vol.~14, no.~3, 2018.
	
	\bibitem[Guti\'{e}rrez-Maldonado et~al.(2016)Guti\'{e}rrez-Maldonado,
	Ferrer-Garc\'{i}a, Pla-Sanjuanelo, Andres-Pueyo, and Caparros]{maldonado2016}
	J.~Guti\'{e}rrez-Maldonado, M.~Ferrer-Garc\'{i}a, J.~Pla-Sanjuanelo,
	A.~Andres-Pueyo, and T.~Caparros, ``Virtual reality to train diagnostic
	skills in eating disorders. comparison of two low cost systems,''
	\emph{Studies in health technology and informatics}, vol. 219, pp. 75--81,
	2016.
	
	\bibitem[Preukschas et~al.(2024)Preukschas, Wise, Bettscheider, Pfeiffer,
	Wagner, Huber, Golriz, Fischer, Mehrabi, R\"{o}ssler, Speidel, Hacker,
	M\"{u}ller-Stich, Nickel, and Kenngott]{preukschas2024}
	A.~A. Preukschas, P.~A. Wise, L.~Bettscheider, M.~Pfeiffer, M.~Wagner,
	M.~Huber, M.~Golriz, L.~Fischer, A.~Mehrabi, F.~R\"{o}ssler, S.~Speidel,
	T.~Hacker, B.~P. M\"{u}ller-Stich, F.~Nickel, and H.~G. Kenngott, ``Comparing
	a virtual reality head-mounted display to on-screen three-dimensional
	visualization and two-dimensional computed tomography data for training in
	decision making in hepatic surgery: a randomized controlled study,''
	\emph{Surg Endosc}, vol.~38, pp. 2483--2496, 2024.
	
	\bibitem[Smith et~al.(2018)Smith, Farra, Ulrich, Hodgson, Nicely, and
	Mickle]{smith2018}
	S.~J. Smith, S.~L. Farra, D.~L. Ulrich, E.~Hodgson, S.~Nicely, and A.~Mickle,
	``Effectiveness of two varying levels of virtual reality simulation,''
	\emph{Nursing education perspectives}, vol.~39, no.~6, 2018.
	
	\bibitem[Rai et~al.(2019)Rai, Tan, and Leo]{rai2019}
	B.~Rai, H.~S. Tan, and C.~H. Leo, ``Bringing play back into the biology
	classroom with the use of gamified virtual lab simulations,'' \emph{Journal
		of Applied Learning \& Teaching}, 2019.
	
	\bibitem[Makransky and Lilleholt(2018)]{makransky2018}
	G.~Makransky and L.~Lilleholt, ``A structural equation modeling investigation
	of the emotional value of immersive virtual reality in education,''
	\emph{Education Tech Research Dev}, vol.~66, pp. 1141--1164, 2018.
	
	\bibitem[Klingenberg et~al.(2020)Klingenberg, Jørgensen, Dandanell, Skriver,
	Mottelson, and Makransky]{klingenberg2020}
	S.~Klingenberg, M.~L.~M. Jørgensen, G.~Dandanell, K.~Skriver, A.~Mottelson,
	and G.~Makransky, ``Investigating the effect of teaching as a generative
	learning strategy when learning through desktop and immersive vr: A media and
	methods experiment,'' \emph{British Journal of Educational Technology},
	vol.~51, no.~6, pp. 2115--2138, 2020.
	
	\bibitem[Kozhevnikov et~al.(2013)Kozhevnikov, Gurlitt, and
	Kozhevnikov]{kozhevnikov2013}
	M.~Kozhevnikov, J.~Gurlitt, and M.~Kozhevnikov, ``Learning relative motion
	concepts in immersive and non-immersive virtual environments,'' \emph{J Sci
		Educ Technol}, vol.~22, pp. 952--962, 2013.
	
	\bibitem[Parmar et~al.(2023)Parmar, Lin, DSouza, Jörg, Leonard, Daily, and
	Babu]{parmar2023}
	D.~Parmar, L.~Lin, N.~DSouza, S.~Jörg, A.~E. Leonard, S.~B. Daily, and S.~V.
	Babu, ``How immersion and self-avatars in vr affect learning programming and
	computational thinking in middle school education,'' \emph{IEEE Transactions
		on Visualization and Computer Graphics}, vol.~29, no.~8, pp. 3698--3713,
	2023.
	
	\bibitem[Oberd\"{o}rfer and Latoschik(2019)]{oberdorfer2019}
	S.~Oberd\"{o}rfer and M.~E. Latoschik, ``Knowledge encoding in game mechanics:
	Transfer-oriented knowledge learning in desktop-3d and vr,''
	\emph{International Journal of Computer Games Technology}, vol. 2019, no.~1,
	p. 7626349, 2019.
	
	\bibitem[Loup-Escande et~al.(2017)Loup-Escande, Jamet, Ragot, Erhel, and
	and]{loup-escande2017}
	E.~Loup-Escande, E.~Jamet, M.~Ragot, S.~Erhel, and N.~M. and, ``Effects of
	stereoscopic display on learning and user experience in an educational
	virtual environment,'' \emph{International Journal of Human--Computer
		Interaction}, vol.~33, no.~2, pp. 115--122, 2017.
	
	\bibitem[Murcia-L\'{o}pez and Steed(2016)]{murcia-lopez2016}
	M.~Murcia-L\'{o}pez and A.~Steed, ``The effect of environmental features,
	self-avatar, and immersion on object location memory in virtual
	environments,'' \emph{Frontiers in ICT}, vol.~3, 2016.
	
	\bibitem[Krokos et~al.(2019)Krokos, Plaisant, and Varshney]{krokos2019}
	E.~Krokos, C.~Plaisant, and A.~Varshney, ``Virtual memory palaces: immersion
	aids recall,'' \emph{Virtual Reality}, vol.~23, pp. 1--15, 2019.
	
	\bibitem[Hejtmanek et~al.(2020)Hejtmanek, Starrett, Ferrer, and
	Ekstrom]{hejtmanek2020}
	L.~Hejtmanek, M.~Starrett, E.~Ferrer, and A.~D. Ekstrom, ``How much of what we
	learn in virtual reality transfers to real-world navigation?''
	\emph{Multisensory research}, vol.~33, pp. 479--503, 2020.
	
	\bibitem[Zhao et~al.(2020)Zhao, Sensibaugh, Bodenheimer, McNamara, Nazareth,
	Newcombe, Minear, and Klippel]{zhao2020}
	J.~Zhao, T.~Sensibaugh, B.~Bodenheimer, T.~P. McNamara, A.~Nazareth,
	N.~Newcombe, M.~Minear, and A.~Klippel, ``Desktop versus immersive virtual
	environments: effects on spatial learning,'' \emph{Spatial Cognition \&
		Computation}, vol.~20, no.~4, pp. 328--363, 2020.
	
	\bibitem[Ochs and Sonderegger(2022)]{ochs2022}
	C.~Ochs and A.~Sonderegger, ``The interplay between presence and learning,''
	\emph{Frontiers in Virtual Reality}, vol.~3, 2022.
	
	\bibitem[Olmos~Raya et~al.(2018)Olmos~Raya, Cavalcanti, Contero, Castellanos,
	Alice, Chicchi~Giglioli, and Alcañiz~Raya]{raya2018}
	E.~Olmos~Raya, J.~Cavalcanti, M.~Contero, M.~Castellanos, I.~Alice,
	I.~Chicchi~Giglioli, and M.~Alcañiz~Raya, ``Mobile virtual reality as an
	educational platform: A pilot study on the impact of immersion and positive
	emotion induction in the learning process,'' \emph{Eurasia Journal of
		Mathematics, Science and Technology Education}, vol.~14, 2018.
	
	\bibitem[Johnson et~al.(2020)Johnson, Damian, and Tzanetakis]{johnson2020}
	D.~Johnson, D.~Damian, and G.~Tzanetakis, ``Evaluating the effectiveness of
	mixed reality music instrument learning with the theremin,'' \emph{Virtual
		Reality}, vol.~24, pp. 303--317, 2020.
	
	\bibitem[Clerici et~al.(2022)Clerici, Boffi, Lanzi, Coppola, Murone, and
	Gallace]{clerici2022}
	M.~Clerici, P.~Boffi, P.~L. Lanzi, L.~Coppola, C.~Murone, and A.~Gallace, ``One
	day in a roman domus: Human factors and educational properties involved in a
	virtual heritage application,'' in \emph{2022 IEEE International Symposium on
		Mixed and Augmented Reality Adjunct (ISMAR-Adjunct)}, 2022, pp. 692--697.
	
	\bibitem[Boffi et~al.(2023{\natexlab{a}})Boffi, Clerici, Gallace, and
	Lanzi]{boffi2023domus}
	P.~Boffi, M.~Clerici, A.~Gallace, and P.~L. Lanzi, ``An educational experience
	in ancient rome to evaluate the impact of virtual reality on human learning
	processes,'' \emph{Computers \& Education: X Reality}, vol.~2, p. 100014,
	2023.
	
	\bibitem[Boffi et~al.(2023{\natexlab{b}})Boffi, Clerici, Muolo, Gallace, and
	Lanzi]{boffi2023ec}
	P.~Boffi, M.~Clerici, M.~Muolo, A.~Gallace, and P.~L. Lanzi, ``Envisioning
	corals: Embodying coral reef inhabitants to raise awareness on climate
	changes impacts on remote environments,'' in \emph{2023 9th International
		Conference on Virtual Reality (ICVR)}, 2023, pp. 239--246.
	
	\bibitem[Battipede et~al.(2025)Battipede, Giangualano, Boffi, Clerici, Calvi,
	Cassenti, Cialini, Van Den~Weghe, Addimando, Lanzi, and
	Gallace]{battipede2025}
	E.~Battipede, A.~Giangualano, P.~Boffi, M.~Clerici, A.~Calvi, L.~Cassenti,
	R.~Cialini, T.~L.~A. Van Den~Weghe, L.~Addimando, P.~L. Lanzi, and
	A.~Gallace, ``Physics playground: Insights from a qualitative-quantitative
	study about vr-based learning,'' in \emph{Intelligent Human Computer
		Interaction}, D.~Singh, J.-W. van't Klooster, and U.~S. Tiwary, Eds., 2025,
	pp. 97--108.
	
	\bibitem[Giangualano et~al.(2025)Giangualano, Boffi, Osimo, Yavari, Gabbiadini,
	Lanzi, and Gallace]{giangualano2025}
	A.~Giangualano, P.~Boffi, S.~A. Osimo, M.~Yavari, A.~Gabbiadini, P.~L. Lanzi,
	and A.~Gallace, ``Embodiment in {Nature}: {How} {Avatar} {Choice} {Shapes} an
	{Underwater} {Virtual} {Reality} {Experience},'' in \emph{2025 {IEEE}
		{International} {Conference} on {Metrology} for {eXtended} {Reality},
		{Artificial} {Intelligence} and {Neural} {Engineering} ({MetroXRAINE})},
	2025, pp. 506--511.
	
	\bibitem[Clifton and Palmisano(2020)]{clifton2020}
	J.~Clifton and S.~Palmisano, ``Effects of steering locomotion and teleporting
	on cybersickness and presence in hmd-based virtual reality,'' \emph{Virtual
		Reality}, vol.~24, pp. 453--468, 2020.
	
	\bibitem[Davis et~al.(1989)Davis, Bagozzi, and Warshaw]{davis1989}
	F.~D. Davis, R.~P. Bagozzi, and P.~R. Warshaw, ``User acceptance of computer
	technology: A comparison of two theoretical models,'' \emph{Management
		Science}, vol.~35, no.~8, pp. 982--1003, 1989.
	
	\bibitem[Rese et~al.(2017)Rese, Baier, Geyer-Schulz, and Schreiber]{rese2017}
	A.~Rese, D.~Baier, A.~Geyer-Schulz, and S.~Schreiber, ``How augmented reality
	apps are accepted by consumers: A comparative analysis using scales and
	opinions,'' \emph{Technological Forecasting and Social Change}, vol. 124, pp.
	306--319, 2017.
	
	\bibitem[Salloum et~al.(2019)Salloum, Qasim Mohammad~Alhamad, Al-Emran,
	Abdel~Monem, and Shaalan]{salloum2019}
	S.~A. Salloum, A.~Qasim Mohammad~Alhamad, M.~Al-Emran, A.~Abdel~Monem, and
	K.~Shaalan, ``Exploring students' acceptance of e-learning through the
	development of a comprehensive technology acceptance model,'' \emph{IEEE
		Access}, vol.~7, pp. 128\,445--128\,462, 2019.
	
	\bibitem[Hassenzahl(2004)]{hassenzahl2004}
	M.~Hassenzahl, \emph{The Thing and I: Understanding the Relationship Between
		User and Product}.\hskip 1em plus 0.5em minus 0.4em\relax Springer
	Netherlands, 2004, pp. 31--42.
	
	\bibitem[Sagnier et~al.(2020)Sagnier, Loup-Escande, Lourdeaux, Thouvenin, and
	Vall\'{e}ry]{sagnier2020}
	C.~Sagnier, E.~Loup-Escande, D.~Lourdeaux, I.~Thouvenin, and G.~Vall\'{e}ry,
	``User acceptance of virtual reality: An extended technology acceptance
	model,'' \emph{International Journal of Human--Computer Interaction},
	vol.~36, no.~11, pp. 993--1007, 2020.
	
	\bibitem[Laugwitz et~al.(2008)Laugwitz, Held, and Schrepp]{ueq2008}
	B.~Laugwitz, T.~Held, and M.~Schrepp, ``Construction and evaluation of a user
	experience questionnaire,'' in \emph{HCI and Usability for Education and
		Work}, 2008, pp. 63--76.
	
	\bibitem[Schrepp et~al.(2017)Schrepp, Hinderks, and Thomaschewski]{ueq-s2017}
	M.~Schrepp, A.~Hinderks, and J.~Thomaschewski, ``Design and evaluation of a
	short version of the user experience questionnaire (ueq-s),''
	\emph{International Journal of Interactive Multimedia and Artificial
		Intelligence}, vol.~4, pp. 103--108, 2017.
	
	\bibitem[Gonzalez-Franco and Peck(2018)]{gonzalez-franco2018}
	M.~Gonzalez-Franco and T.~C. Peck, ``Avatar embodiment. towards a standardized
	questionnaire,'' \emph{Frontiers in Robotics and AI}, vol.~5, 2018.
	
	\bibitem[Kim et~al.(2018)Kim, Park, Choi, and Choe]{kim2018}
	H.~K. Kim, J.~Park, Y.~Choi, and M.~Choe, ``Virtual reality sickness
	questionnaire (vrsq): Motion sickness measurement index in a virtual reality
	environment,'' \emph{Applied Ergonomics}, vol.~69, pp. 66--73, 2018.
	
	\bibitem[Hart and Staveland(1988)]{tlx-nasa}
	S.~G. Hart and L.~E. Staveland, ``Development of nasa-tlx (task load index):
	Results of empirical and theoretical research,'' in \emph{Human Mental
		Workload}, 1988, vol.~52, pp. 139--183.
	
	\bibitem[Brooke(1995)]{sus}
	J.~Brooke, ``Sus: A quick and dirty usability scale,'' \emph{Usability Eval.
		Ind.}, vol. 189, 1995, [Online]. Available
	\href{https://tinyurl.com/sus-source}{\textbf{here}}.
	
	\bibitem[Kyrlitsias et~al.(2020)Kyrlitsias, Christofi, Michael-Grigoriou,
	Banakou, and Ioannou]{kyrlitsias2020}
	C.~Kyrlitsias, M.~Christofi, D.~Michael-Grigoriou, D.~Banakou, and A.~Ioannou,
	``A virtual tour of a hardly accessible archaeological site: The effect of
	immersive virtual reality on user experience, learning and attitude change,''
	\emph{Frontiers in Computer Science}, vol.~2, 2020.
	
	\bibitem[Scurati et~al.(2021)Scurati, Bertoni, Graziosi, and
	Ferrise]{scurati2021}
	G.~W. Scurati, M.~Bertoni, S.~Graziosi, and F.~Ferrise, ``Exploring the use of
	virtual reality to support environmentally sustainable behavior: A framework
	to design experiences,'' \emph{Sustainability}, vol.~13, no.~2, 2021.
	
	\bibitem[Loorbach et~al.(2015)Loorbach, Peters, Karreman, and
	Steehouder]{loorbach2015}
	N.~Loorbach, O.~Peters, J.~Karreman, and M.~Steehouder, ``Validation of the
	instructional materials motivation survey (imms) in a self-directed
	instructional setting aimed at working with technology,'' \emph{British
		Journal of Educational Technology}, vol.~46, no.~1, pp. 204--218, 2015.
	
	\bibitem[Perkins et~al.(2006)Perkins, Adams, Dubson, Finkelstein, Reid, Wieman,
	and LeMaster]{perkins2006}
	K.~Perkins, W.~Adams, M.~Dubson, N.~Finkelstein, S.~Reid, C.~Wieman, and
	R.~LeMaster, ``Phet: Interactive simulations for teaching and learning
	physics,'' \emph{The Physics Teacher}, vol.~44, no.~1, pp. 18--23, 2006.
	
	\bibitem[Hamilton et~al.(2021)Hamilton, McKechnie, Edgerton, and
	Wilson]{hamilton2021}
	D.~Hamilton, J.~McKechnie, E.~Edgerton, and C.~Wilson, ``Immersive virtual
	reality as a pedagogical tool in education: a systematic literature review of
	quantitative learning outcomes and experimental design,'' \emph{Journal of
		Computers in Education}, vol.~8, no.~1, 2021.
	
	\bibitem[Lai et~al.(2022)Lai, Lin, Chou, and Yueh]{lai2022}
	T.-L. Lai, Y.-S. Lin, C.-Y. Chou, and H.-P. Yueh, ``Evaluation of an
	inquiry-based virtual lab for junior high school science classes,''
	\emph{Journal of Educational Computing Research}, vol.~59, no.~8, pp.
	1579--1600, 2022.
	
	\bibitem[Makransky et~al.(2019{\natexlab{b}})Makransky, Borre-Gude, and
	Mayer]{makransky2019a}
	G.~Makransky, S.~Borre-Gude, and R.~E. Mayer, ``Motivational and cognitive
	benefits of training in immersive virtual reality based on multiple
	assessments,'' \emph{Journal of Computer Assisted Learning}, vol.~35, no.~6,
	pp. 691--707, 2019.
	
	\bibitem[Ding et~al.(2024)Ding, Huang, DuBois, and Fu]{ding2024}
	A.-C.~E. Ding, K.-T.~T. Huang, J.~DuBois, and H.~Fu, ``Integrating immersive
	virtual reality technology in scaffolded game-based learning to enhance low
	motivation students' multimodal science learning,'' \emph{Education Tech
		Research Dev}, vol.~72, pp. 2083--2102, 2024.
	
	\bibitem[Ortiz and Elizondo(2023)]{ortiz2023}
	A.~Ortiz and S.~Elizondo, ``Design of an immersive virtual reality framework to
	enhance the sense of agency using affective computing technologies,''
	\emph{Applied Sciences}, vol.~13, no.~24, 2023.
	
	\bibitem[McGivney(2025)]{mcgivney2025}
	E.~McGivney, ``Interactivity and identity impact learners' sense of agency in
	virtual reality field trips,'' \emph{British Journal of Educational
		Technology}, vol.~56, no.~1, pp. 410--434, 2025.
	
	\bibitem[K\"{u}hnel et~al.(2023)K\"{u}hnel, Kecelioglu, Maltby, Hood, Knott,
	Ditton, Walker, and Kluge]{kuhnel2023}
	C.~K\"{u}hnel, E.~D. Kecelioglu, S.~Maltby, R.~Hood, B.~Knott, E.~Ditton,
	F.~Walker, and M.~Kluge, ``Direct comparison of virtual reality and 2d
	delivery on sense of presence, emotional and physiological outcome
	measures,'' \emph{Frontiers Virtual Real.}, vol.~4, 2023.
	
	\bibitem[Niu et~al.(2019)Niu, Wang, Wang, Sun, Yue, and Zheng]{niu2019}
	Y.~Niu, D.~Wang, Z.~Wang, F.~Sun, K.~Yue, and N.~Zheng, ``User experience
	evaluation in virtual reality based on subjective feelings and physiological
	signals,'' \emph{Journal of Imaging Science and Technology}, 2019.
	
	\bibitem[Beck and Egger(2018)]{beck2018}
	J.~Beck and R.~Egger, ``Emotionalise me: Self-reporting and arousal
	measurements in virtual tourism environments,'' in \emph{Information and
		Communication Technologies in Tourism 2018}, 2018, pp. 3--15.
	
	\bibitem[Huang(2003)]{huang2003}
	M.-H. Huang, ``Designing website attributes to induce experiential
	encounters,'' \emph{Computers in Human Behavior}, vol.~19, no.~4, pp.
	425--442, 2003.
	
	\bibitem[Fiorella and Mayer(2015)]{GLS_FiorellaMayer2015}
	L.~Fiorella and R.~E. Mayer, \emph{Learning as a Generative Activity: Eight
		Learning Strategies that Promote Understanding}.\hskip 1em plus 0.5em minus
	0.4em\relax Cambridge University Press, 2015.
	
	\bibitem[Tauro et~al.(2022)Tauro, Gorini, Caglio, Gabanelli, and
	Caiani]{tauro2022personas}
	E.~Tauro, A.~Gorini, C.~Caglio, P.~Gabanelli, and E.~G. Caiani, ``Covid-19 and
	mental disorders in healthcare personnel: A novel framework to develop
	personas from an online survey,'' \emph{Journal of Biomedical Informatics},
	vol. 126, p. 103993, 2022.
	
	\bibitem[Sousa and Belchior-Rocha(2024)]{souza2024}
	M.~Sousa and H.~Belchior-Rocha, ``Innovative teaching approaches and
	personalized learning experiences using artificial intelligence,'' in
	\emph{EDULEARN24 Proceedings}, ser. 16th International Conference on
	Education and New Learning Technologies, 2024, pp. 9003--9009.
	
\end{thebibliography}
\end{document}